\documentclass[superscriptaddress,showpacs,amssymb,amsmath,
amsfonts,aps,
prc,twocolumn
]{revtex4}
\usepackage{graphicx}
\usepackage{dcolumn}
\usepackage{epsfig}
\usepackage{subfigure}

\begin{document}

\title{Cross sections  for the 
$\gamma p \to K^{*+} \Lambda$ and 
$\gamma p \to K^{*+} \Sigma^{0}$ reactions measured at CLAS}
%

\newcommand*{\OHIOU}{Ohio University, Athens, Ohio  45701}
\newcommand*{\OHIOUindex}{1}
\affiliation{\OHIOU}
\newcommand*{\ANL}{Argonne National Laboratory, Argonne, Illinois 60439}
\newcommand*{\ANLindex}{2}
\affiliation{\ANL}
\newcommand*{\ASU}{Arizona State University, Tempe, Arizona 85287-1504}
\newcommand*{\ASUindex}{3}
\affiliation{\ASU}
\newcommand*{\CSUDH}{California State University, Dominguez Hills, Carson, CA 90747}
\newcommand*{\CSUDHindex}{4}
\affiliation{\CSUDH}
\newcommand*{\CANISIUS}{Canisius College, Buffalo, NY}
\newcommand*{\CANISIUSindex}{5}
\affiliation{\CANISIUS}
\newcommand*{\CMU}{Carnegie Mellon University, Pittsburgh, Pennsylvania 15213}
\newcommand*{\CMUindex}{6}
\affiliation{\CMU}
\newcommand*{\CUA}{Catholic University of America, Washington, D.C. 20064}
\newcommand*{\CUAindex}{7}
\affiliation{\CUA}
\newcommand*{\SACLAY}{CEA, Centre de Saclay, Irfu/Service de Physique Nucl\'eaire, 91191 Gif-sur-Yvette, France}
\newcommand*{\SACLAYindex}{8}
\affiliation{\SACLAY}
\newcommand*{\CNU}{Christopher Newport University, Newport News, Virginia 23606}
\newcommand*{\CNUindex}{9}
\affiliation{\CNU}
\newcommand*{\UCONN}{University of Connecticut, Storrs, Connecticut 06269}
\newcommand*{\UCONNindex}{10}
\affiliation{\UCONN}
\newcommand*{\EDINBURGH}{Edinburgh University, Edinburgh EH9 3JZ, United Kingdom}
\newcommand*{\EDINBURGHindex}{11}
\affiliation{\EDINBURGH}
\newcommand*{\FU}{Fairfield University, Fairfield CT 06824}
\newcommand*{\FUindex}{12}
\affiliation{\FU}
\newcommand*{\FIU}{Florida International University, Miami, Florida 33199}
\newcommand*{\FIUindex}{13}
\affiliation{\FIU}
\newcommand*{\FSU}{Florida State University, Tallahassee, Florida 32306}
\newcommand*{\FSUindex}{14}
\affiliation{\FSU}
\newcommand*{\GWUI}{The George Washington University, Washington, DC 20052}
\newcommand*{\GWUIindex}{15}
\affiliation{\GWUI}
\newcommand*{\ISU}{Idaho State University, Pocatello, Idaho 83209}
\newcommand*{\ISUindex}{16}
\affiliation{\ISU}
\newcommand*{\INFNFE}{INFN, Sezione di Ferrara, 44100 Ferrara, Italy}
\newcommand*{\INFNFEindex}{17}
\affiliation{\INFNFE}
\newcommand*{\INFNFR}{INFN, Laboratori Nazionali di Frascati, 00044 Frascati, Italy}
\newcommand*{\INFNFRindex}{18}
\affiliation{\INFNFR}
\newcommand*{\INFNGE}{INFN, Sezione di Genova, 16146 Genova, Italy}
\newcommand*{\INFNGEindex}{19}
\affiliation{\INFNGE}
\newcommand*{\INFNRO}{INFN, Sezione di Roma Tor Vergata, 00133 Rome, Italy}
\newcommand*{\INFNROindex}{20}
\affiliation{\INFNRO}
\newcommand*{\ORSAY}{Institut de Physique Nucl\'eaire ORSAY, Orsay, France}
\newcommand*{\ORSAYindex}{21}
\affiliation{\ORSAY}
\newcommand*{\ITEP}{Institute of Theoretical and Experimental Physics, Moscow, 117259, Russia}
\newcommand*{\ITEPindex}{22}
\affiliation{\ITEP}
\newcommand*{\JMU}{James Madison University, Harrisonburg, Virginia 22807}
\newcommand*{\JMUindex}{23}
\affiliation{\JMU}
\newcommand*{\KNU}{Kyungpook National University, Daegu 702-701, Republic of Korea}
\newcommand*{\KNUindex}{24}
\affiliation{\KNU}
\newcommand*{\LPSC}{LPSC, Universite Joseph Fourier, CNRS/IN2P3, INPG, Grenoble, France
}
\newcommand*{\LPSCindex}{25}
\affiliation{\LPSC}
\newcommand*{\UNH}{University of New Hampshire, Durham, New Hampshire 03824-3568}
\newcommand*{\UNHindex}{26}
\affiliation{\UNH}
\newcommand*{\NSU}{Norfolk State University, Norfolk, Virginia 23504}
\newcommand*{\NSUindex}{27}
\affiliation{\NSU}
\newcommand*{\ODU}{Old Dominion University, Norfolk, Virginia 23529}
\newcommand*{\ODUindex}{28}
\affiliation{\ODU}
\newcommand*{\RPI}{Rensselaer Polytechnic Institute, Troy, New York 12180-3590}
\newcommand*{\RPIindex}{29}
\affiliation{\RPI}
\newcommand*{\URICH}{University of Richmond, Richmond, Virginia 23173}
\newcommand*{\URICHindex}{30}
\affiliation{\URICH}
\newcommand*{\ROMAII}{Universita' di Roma Tor Vergata, 00133 Rome Italy}
\newcommand*{\ROMAIIindex}{31}
\affiliation{\ROMAII}
\newcommand*{\MSU}{Skobeltsyn Nuclear Physics Institute, 119899 Moscow, Russia}
\newcommand*{\MSUindex}{32}
\affiliation{\MSU}
\newcommand*{\SCAROLINA}{University of South Carolina, Columbia, South Carolina 29208}
\newcommand*{\SCAROLINAindex}{33}
\affiliation{\SCAROLINA}
\newcommand*{\JLAB}{Thomas Jefferson National Accelerator Facility, Newport News, Virginia 23606}
\newcommand*{\JLABindex}{34}
\affiliation{\JLAB}
\newcommand*{\UTFSM}{Universidad T\'{e}cnica Federico Santa Mar\'{i}a, Casilla 110-V Valpara\'{i}so, Chile}
\newcommand*{\UTFSMindex}{35}
\affiliation{\UTFSM}
\newcommand*{\GLASGOW}{University of Glasgow, Glasgow G12 8QQ, United Kingdom}
\newcommand*{\GLASGOWindex}{36}
\affiliation{\GLASGOW}
\newcommand*{\VIRGINIA}{University of Virginia, Charlottesville, Virginia 22901}
\newcommand*{\VIRGINIAindex}{37}
\affiliation{\VIRGINIA}
\newcommand*{\WM}{College of William and Mary, Williamsburg, Virginia 23187-8795}
\newcommand*{\WMindex}{38}
\affiliation{\WM}
\newcommand*{\YEREVAN}{Yerevan Physics Institute, 375036 Yerevan, Armenia}
\newcommand*{\YEREVANindex}{39}
\affiliation{\YEREVAN}

\newcommand*{\KINHA}{Inha University, Incheon 402-751, Republic of Korea}
\newcommand*{\KINHAindex}{40}
\affiliation{\KINHA}

\newcommand*{\NOWOSAKA}{Osaka University, 567-0047 Ibarakishi, Japan}
\newcommand*{\NOWINFNGE}{INFN, Sezione di Genova, 16146 Genova, Italy}
\newcommand*{\NOWMSU}{Skobeltsyn Nuclear Physics Institute, 119899 Moscow, Russia}
\newcommand*{\NOWORSAY}{Institut de Physique Nucl\'eaire ORSAY, Orsay, France}
\newcommand*{\NOWODU}{Old Dominion University, Norfolk, Virginia 23529}
\newcommand*{\NOWROMAII}{Universita' di Roma Tor Vergata, 00133 Rome Italy}
\newcommand*{\NOWUCONN}{University of Connecticut, Storrs, Connecticut 06269}
\newcommand*{\NOWVIRGINIA}{University of Virginia, Charlottesville, Virginia 22901}

\author {W.~Tang}
\affiliation{\OHIOU}
\author {K.~Hicks}
\affiliation{\OHIOU}
\author {D.~Keller}
\altaffiliation[Current address:] {\NOWVIRGINIA}
\affiliation{\OHIOU}
\author {S.~H.~Kim}
\altaffiliation[Current address:] {\NOWOSAKA}
\affiliation{\KINHA}
\author {H.~C.~Kim}
\affiliation{\KINHA}
\author {K.P. ~Adhikari}
\affiliation{\ODU}
\author {M.~Aghasyan}
\affiliation{\INFNFR}
\author {M.J.~Amaryan}
\affiliation{\ODU}
\author {M.D.~Anderson}
\affiliation{\GLASGOW}
\author {S. ~Anefalos~Pereira}
\affiliation{\INFNFR}
% ...

\collaboration{The CLAS Collaboration}
\noaffiliation
\date{\today}

\begin{abstract}

The first high-statistics cross sections for the reactions 
$\gamma p \to K^{*+} \Lambda$ and $\gamma p \to K^{*+} \Sigma^0$ 
were measured using the CLAS detector at photon energies between 
threshold and 3.9 GeV at the Thomas Jefferson National Accelerator 
Facility. Differential cross sections are presented over the 
full range of the center-of-mass angles, $\theta^{CM}_{K^{*+}}$, and 
then fitted to Legendre polynomials to extract the total cross 
section.  Results for the $K^{*+}\Lambda$ final state are compared 
with two different calculations in an isobar and a Regge model, respectively. 
Theoretical calculations significantly underestimate the $K^{*+} \Lambda$ 
total cross sections between 2.1 and 2.6 GeV, but are in better 
agreement with present data at higher photon energies.
\end{abstract}

\maketitle

\section{Introduction}
\hspace{5mm}
\linespread{1.6}
One motivation for the study of $K^*$ photoproduction is to 
investigate the role of the $K^*_0(800)$ meson (also called the $\kappa$) 
through $t$-channel exchange. The $\kappa$ 
is expected to be in the same scalar meson nonet as 
the $f_0(500)$ meson (also called the $\sigma$).  Neither of these 
mesons have been directly observed because of their large widths, 
which are nearly as big as their respective masses.  Such a 
large width is expected for scalar mesons, which have quantum 
numbers $J^{PC} = 0^{++}$. In many quark models, there is 
virtually no angular momentum barrier to
prevent these mesons from falling apart into two mesons, such 
as $\sigma \to \pi \pi$ or $\kappa \to K \pi$.  Because the $\sigma$ 
and $\kappa$ mesons cannot be observed directly, indirect production 
mechanisms provide better evidence of their existence. 

The $\sigma$ meson is rather well established \cite{pdg} 
as a $\pi \pi$ resonance, which is an important component of 
models of the nucleon-nucleon ($N$-$N$) interaction such as 
the Bonn potential \cite{cdbonn}. The $\kappa$ meson, however, 
is less easily established due to its strange quark content. 
Data for hyperon-nucleon ($Y$-$N$) interactions are sparse 
and hence models have a range of parameter space that may 
or may not include $\kappa$ exchange. Perhaps the best current 
evidence for the $\kappa$ is from the decay angular distributions 
 of the $D$-meson into $K\pi\pi$ final states \cite{belle}. 

Here, in photoproduction of the $K^{*+}$, the $\kappa^+$ enters 
into the $t$-channel exchange diagrams \cite{oh2}.  $\kappa$ cannot 
contribute to kaon photoproduction because the photon cannot 
couple to the $K$-$\kappa$ vertex due to G-parity conservation. 
Theoretical calculations have been done \cite{oh2} showing the 
effect of the $\kappa$ on photoproduction of $K^{*+}$ and 
$K^{*0}$ final states. Several years ago, two reports of 
$K^{*0}$ photoproduction were published \cite{hleiqawi,cbelsa} 
but only preliminary results on $K^{*+}$ were available \cite{guo}.  

We present the first results of 
$K^{*+}\Lambda$  and $K^{*+}\Sigma^0$ photoproduction with 
high statistics.  Together, the $K^{*+}$ and $K^{*0}$ 
photoproduction results could put significant constraints on 
the role of the $\kappa$ meson in $t$-channel exchange.
Here, for the first time, we make the ratio of total cross sections for the 
reactions $\gamma p \to K^{*+}\Lambda$ and $\gamma p \to K^{*0}\Sigma^+$ 
and compare with the same ratio calculated from a theoretical 
model for large and small contributions from $\kappa$ exchange.
Other evidence for the $\kappa$ comes from recently published data on 
the linear beam asymmetry in photoproduction of the 
$\vec{\gamma} p \to K^{*0}\Sigma^+$ reaction \cite{hwang} 
which shows a significant positive value at forward $K^*$ angles 
that is the signature of $\kappa$ exchange \cite{oh2}.

A secondary motivation for this study is to understand if 
theoretical models using Regge trajectories plus known 
baryon resonances can explain the $K^{*+}$ photoproduction data. 
If not, then there may be higher-mass baryon resonances 
that could couple strongly to $K^*Y$ decay. In a classic 
paper on the quark model, Capstick and Roberts calculated 
\cite{capstick}  
many nucleon resonances that were predicted, but not observed 
in existing partial wave analyses of pion-nucleon scattering. 
They also observed that some of the higher-mass resonances 
may couple weakly to pion decay channels and more strongly 
to $KY$ and $K^*Y$ decays.  Indeed, studies of $KY$ 
photoproduction \cite{bradford2} have shown that hadronic 
model calculations cannot explain the data without the 
addition of a new nucleon resonance near 1.9 GeV. We can look for other "missing resonance states at higher mass, such as those identified in the Bonn-Gatchina analysis \cite{gatchina} largely through precise hyperon photoproduction data from CLAS, 
by comparing  $K^*$ photoproduction data to model calculations.


This paper is organized into the following sections.  First, 
the experiment is described.  Next, the data analysis is 
presented in some detail.  Then we compare the results with 
theoretical calculations.  Finally, we discuss the significance 
of the comparison and provide some conclusions.

\section{Experimental Setup}

The data used in this analysis are from part of the g11a experiment, 
which was taken from May 17 to July 29, 2004, using the CEBAF Large 
Acceptance Spectrometer (CLAS) located in  Hall-B  at the Thomas 
Jefferson National Accelerator Facility (TJNAF) in Newport News, Virginia. 
Real photons were produced by bremsstrahlung from a 4.0186 GeV 
electron beam incident on a 1 $\times$ 10$^{-4}$ radiation length gold foil.
The electron beam was delivered by the Continuous Electron Beam 
Accelerator Facility (CEBAF). 
The Hall-B Tagging System \cite{sober} was used to determine the 
photon energies by measuring the energies of the recoil electrons 
using a dipole magnetic field and a scintillator hodoscope. 
The associated photon energies were then calculated by the difference 
between the incident electron energies and the recoil electron energies 
with an energy resolution of about 2-3 MeV. 
The Hall-B Tagging System tags photons in the range from 20\% to 95\% of the 
incident electron energy.\\ 
\indent A liquid hydrogen target was used in the g11a experiment.  The 
target was contained in a cylindrical Kapton chamber of 2 cm radius and 40 cm length. The target density was determined by the temperature and pressure, which were monitored once per hour during the g11a experiment running.\\
\indent The CLAS apparatus was used to detect particles generated from the 
interaction of the incident photons with the target. 
The CLAS detector was able to track charged particles that have momenta 
larger than $\sim$200 MeV, and the detection area covered polar angles 
from 8$^{\circ}$ to 142$^{\circ}$ and 80\% of the azimuthal region.  
It was composed of several sub-systems, arranged with 
a six-fold azimuthal symmetry. 
A plastic scintillator Start Counter, placed just outside of the target, 
was used to measure the vertex time of particles in coincidence with 
the incoming photon. 
The Start Counter was made of 24 scintillator strips 
with a time resolution of $\sim$ 350 ps \cite{sharabian}. 
The superconducting coils of the CLAS detector generated a toroidal 
magnetic field that bent the path of outgoing charged particles. 
Those particles traveled through 3 regions of drift chambers \cite{mestayer} 
that measured the curved paths to give the particle momenta
with a typical resolution of $\sim$1.0\%. 
For the g11a experiment, the current in the superconducting coils was 
set at 1920 A, which gave a maximum magnetic field of $\sim$ 1.8 T. 
The time of flight (TOF) system was located beyond the outermost drift 
chambers at a radius of $\sim$4 m from the target and was used to measure the time and position of each charged 
particle that hit the TOF scintillators. 
The TOF information, along with the particle momentum, was used for 
the particle identification in the analysis.  
The time resolution of the TOF system was about 80 ps to 160 ps, 
depending on the length of the scintillators \cite{smith}. 
A more detailed description of the CLAS detector is given in 
Ref. \cite{mecking}. \\
\indent The event trigger for the g11a experiment required that at least 
two tracks were detected in different sectors of CLAS. 
Once the event satisfied this condition, it was written to tape for 
future analysis. 
The data acquisition system for the g11a experiment was able to run 
at $\sim$ 5 kHz with a typical livetime of 90\%.

\section{Data Analysis}

As one of the largest photoproduction datasets at CLAS, the g11a experiment 
has $\sim$20 billion triggers.  
The calibration of each CLAS sub-system followed the same 
procedures as described in Ref. \cite{mike}. 
Additional details can be found in Ref. \cite{Tang}.

\subsection{Channels of Interest}

Because the $K^{*+}$ is an unstable particle, it will quickly decay 
to $K\pi$ (see Table \ref{particle}) by the strong interaction.  
By applying energy and momentum conservation, the $K^{*+}$ 
momentum is reconstructed from its decay particles, $K^0 \pi^{+}$. 
The $K^0$ is a mixture of 50\% $K_{S}$ and 50\% $K_{L}$, 
but only the $K_S$ decay is detected by the CLAS detector. 
The $K^{0}$ is reconstructed from the $K_S$ decay to $\pi^+ \pi^-$  
with a decay branching fraction of 69.2\%.
The same branching fractions are reproduced by the Monte Carlo detector
simulations (see section 3.6), and hence are implicit in the detector 
acceptance values.

\begin{table}
\caption{Some physical properties of the $K^{*+}$, $\Lambda$ and $\Sigma^{0}$ \cite{pdg}. }
\begin{center}
\begin{tabular}{lccc}
\hline
\hline
    & $K^{*+}$ & $\Lambda$ & $\Sigma^{0}$ \\ \hline
Mass(GeV)&0.89166 &1.11568&1.19264 \\
Decay products &  $K^0$$\pi^{+}$, $K^+$$\pi^{0}$  & $p\pi^{-}$, $n\pi^{0}$ & $\Lambda$$\gamma$\\  
Branching fraction & 66.7\%, 33.3\% & 63.9\%, 35.8\% &100\% \\ \hline
\end{tabular}
\end{center}
\label{particle}
\end{table} 

To summarize, we report on the differential and total cross sections 
of the photoproduction channel(s):
\begin{equation}
\gamma p \to K^{*+} \Lambda(\Sigma^{0})
\label{eq1}
\end{equation}
followed by
\begin{equation}  
 K^{*+} \to K^{0} \pi^{+}       
\label{eq2}
\end{equation}
and
\begin{equation}  
K_{S} \to  \pi^{+}  \pi^{-} \ .
\label{eq3}
\end{equation}

The $K^{*+}$  and  $K_{S}$ are reconstructed directly from their 
decay products, while  the $\Lambda$ and $\Sigma^{0}$ 
are reconstructed using the missing mass technique.

\subsection{Particle Identification}

\begin{figure}
\begin{center}
\epsfig{file=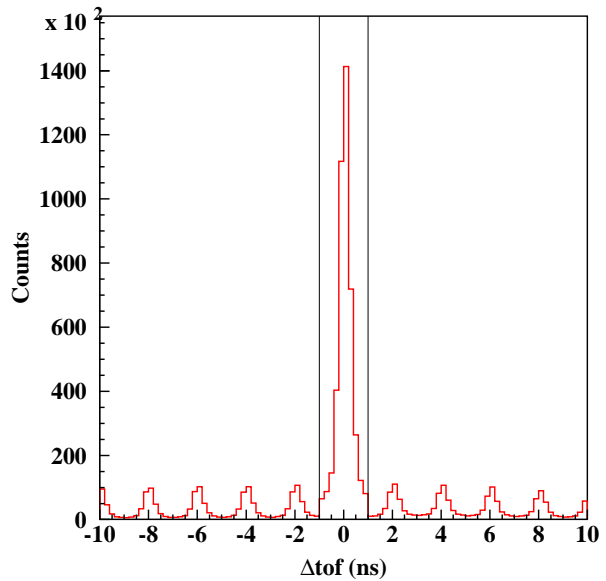,width=\columnwidth}
\caption{ The TOF difference spectrum for pions. The two straight lines show the cut limits for selecting pions in a time window of $\pm$1.0 ns.}
\label{pid}
\end{center}
\end{figure}

The Time of Flight (TOF) difference method was 
used to identify events with three pions (two positive and one negative 
charge) in the final state. Explicitly, 
\begin{equation}
\Delta tof = tof_{mea} - tof_{cal},
\label{eq4}
\end{equation}
where $tof_{mea}$ is the measured TOF of the particle and 
$tof_{cal}$ is the calculated TOF with the measured momentum $p$ 
and the mass of a pion. In more detail, 
\begin{equation}
tof_{mea} = t_{tof} - t_{st},
\label{eq5}
\end{equation}
where $t_{tof}$ is the time when the particle hits the TOF scintillators and 
 $t_{st}$ is the time when the photon hits the target. This information is determined by the CLAS Start Counter. In comparison, $tof_{cal}$ is given by:
\begin{equation}
tof_{cal} = \frac{L}{c} \cdot \frac{1}{\beta}, 
\label{eq6}
\end{equation}
where 
\begin{equation}
\beta = \frac{p}{\sqrt{p{^2} + m{^2}}}.
\label{eq7}
\end{equation}
Thus
\begin{equation}
tof_{cal} = \frac{L}{c} \cdot \sqrt{1 + {\frac{m {^2}}{p{^2}}}}, 
\label{eq8}
\end{equation}
where $L$ is the path length from the target to the TOF scintillators, 
$c$ is the speed of light, $p$ is the particle's momentum, and 
$m$ is the mass of a pion. 
The pion candidates are required to have $|\Delta tof| < 1.0$ ns. 
Fig. \ref{pid} shows the TOF difference spectrum. The solid lines define the region of the cut, the small peaks on the both side of the cuts are due to photons coming from other beam bunches, showing evidence of the $\sim$2 ns beam bunch structure of CEBAF. 

\subsection{Photon Selection}

After applying the  $|$$\Delta tof$$|$ $<$ 1.0  ns cut, particles that came from 
different RF beam buckets were removed naturally. 
Of the photons measured by the photon tagger, we want those
that come within 1.0 ns of the particle vertex time, which are 
called ``good'' photons. However, there might still be more than one ``good'' photon in each event. To select the correct photon, all ``good'' photons were scanned to find the one that gave the three-pion missing mass 
closest to the known mass of the $\Lambda$($\Sigma^{0}$), where 
\begin{equation}
MM(\pi^{+} \pi^{-} \pi^{+}) = \sqrt{ 
\left( E_\gamma + m_p - \sum E_\pi \right)^2 - 
\left( \vec{p}_\gamma - \sum \vec{p}_\pi \right)^2}
\end{equation}
is the missing mass summed over all three pions in the event, 
while $E_\gamma$ and $\vec{p}_\gamma$ are the energy and momentum 
vector of the photon.  The two-pion missing mass is similarly defined.

\subsection{Cuts applied}
\label{cuts}

Several cuts were applied to the data to reduce the background and 
to remove events below threshold for the reaction of interest.
In general, the strategy is to use geometric and kinematic 
constraints to eliminate backgrounds while ensuring that the 
signal remains robust.  The efficiency of various cuts was 
tested with Monte Carlo simulations (see section \ref{da}).


The geometric and kinematic constraints used here are listed below:
\begin{itemize}
\item
Fiducial cuts were applied to remove events that were detected in 
regions of the CLAS detector where the calibration of the detector is 
not well understood.

\item
A cut on the vertex position along the beam axis (the $z$-axis) to be 
within the target position was applied.  All pions were required to 
be generated from the same vertex position within the experimental position uncertainty.

\item
The missing mass from the $K^0$ was required to satisfy the relation $MM(\pi^+ \pi^-) > 1.0$ GeV to include all hyperon mass peaks, 
for pion pairs with an invariant mass inside the $K^0$ mass window 
(see next section).  
Similarly, the missing mass from the $K^{*+}$ was required to be 
greater than the nucleon mass, $MM(\pi^+ \pi^- \pi^+) > 1.0$ GeV.

\noindent After this step, the $K^{*+} \Lambda$ and $K^{*+} \Sigma^0$ reaction channels were treated differently, since different backgrounds are present for each final state. For instance, the large background from 
\begin{equation}
\gamma p \to K^{0} \Sigma{}^{\ast{}+}(1385)
\label{eqxxx}
\end{equation}
present for the  $K^{*+} \Lambda$ reaction channel makes the  the extraction of $K^{*+} \Lambda$ yields by simply fitting the $\Lambda$ peak impossible. On the other hand, there are only very small portions of the $\Sigma^{\ast{}+}$(1385) that contribute to the $K^{*+} \Sigma^0$ background, which can be easily removed based on Monte Carlo studies (see following section). Thus we could fit directly the $\Sigma^0$ peak in the three-pion missing mass for $K^{*+} \Sigma^0$ channel, whereas a different approach (given below) is necessary to extract the $K^{*+} \Lambda$ yield separately from background due to $K^0 \Sigma^{\ast{}+}$(1385) production. The following lists the extra cuts applied for each reaction channel. 

\item 
For the $K^{*+} \Lambda$ analysis, a cut was placed on the 
$\Lambda$ peak in the three-pion missing mass:
1.08 GeV $< MM(\pi^+ \pi^- \pi^+) <$ 1.15 GeV.
This ensures that a $\Lambda$ was present in the final state.

\item 
For the $K^{*+} \Sigma^0$ analysis, a cut was placed on the 
$K^{*+}$ peak of the three-pion invariant mass:
 0.812 GeV $< M(\pi^+ \pi^- \pi^+) <$ 0.972 GeV.
This ensures that a $K^{*+}$ was produced.

\end{itemize}

\subsection{ Sideband Subtraction }
\label{ssb}

\begin{figure}
\epsfig{file=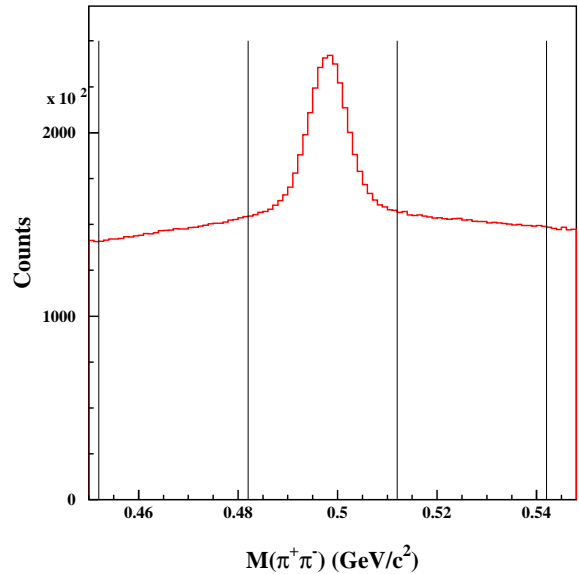,width=\columnwidth}
\caption{The reconstructed two-pion invariant mass showing the $K_{S}$  
distribution. The vertical lines define the bands for the SSM, 
as explained in the text.
}
\label{k0}
\end{figure}

Because reactions other than $K^{*}$ photoproduction are present, 
background is still mixed in with the channels of interest. 
Fig.~\ref{k0} shows the two-pion invariant mass plot after the 
first three cuts in the previous section, integrated over all 
photon energies.  A clear peak centered near 0.497 GeV sits on 
top of a smooth background. 
The invariant mass is calculated using the momentum vector of one 
 $\pi^+$ in the event, along with the $\pi^-$ momentum.
Since there are two $\pi^+$s, both $\pi^+ \pi^-$ pairs are tested, 
but typically only one combination will satisfy all kinematic 
constraints.  To avoid double-counting, in rare cases where 
both $\pi^+$ satisfy all constraints, this combinatoric background 
is removed, for both data analysis and Monte Carlo acceptances.

To reduce the background, a Sideband Subtraction Method (SSM) was applied. 
The concept of the SSM is to assume that the background in the signal 
region can be approximated by a combination of the left and the right 
regions, which are adjacent to the signal region. 
In our analysis, the two-pion mass of the $K_{S}$ is used as the criteria 
to select the signal and sideband regions. 
Fig.~\ref{k0} shows the regions used 
in our analysis. The middle band is the signal 
region, centered at the mass of $K^{0}$ with a width of 0.03 GeV. 
The other two bands, with the same band sizes, are the combinatorial 
background.  

\begin{figure*}[htp]
\epsfig{file=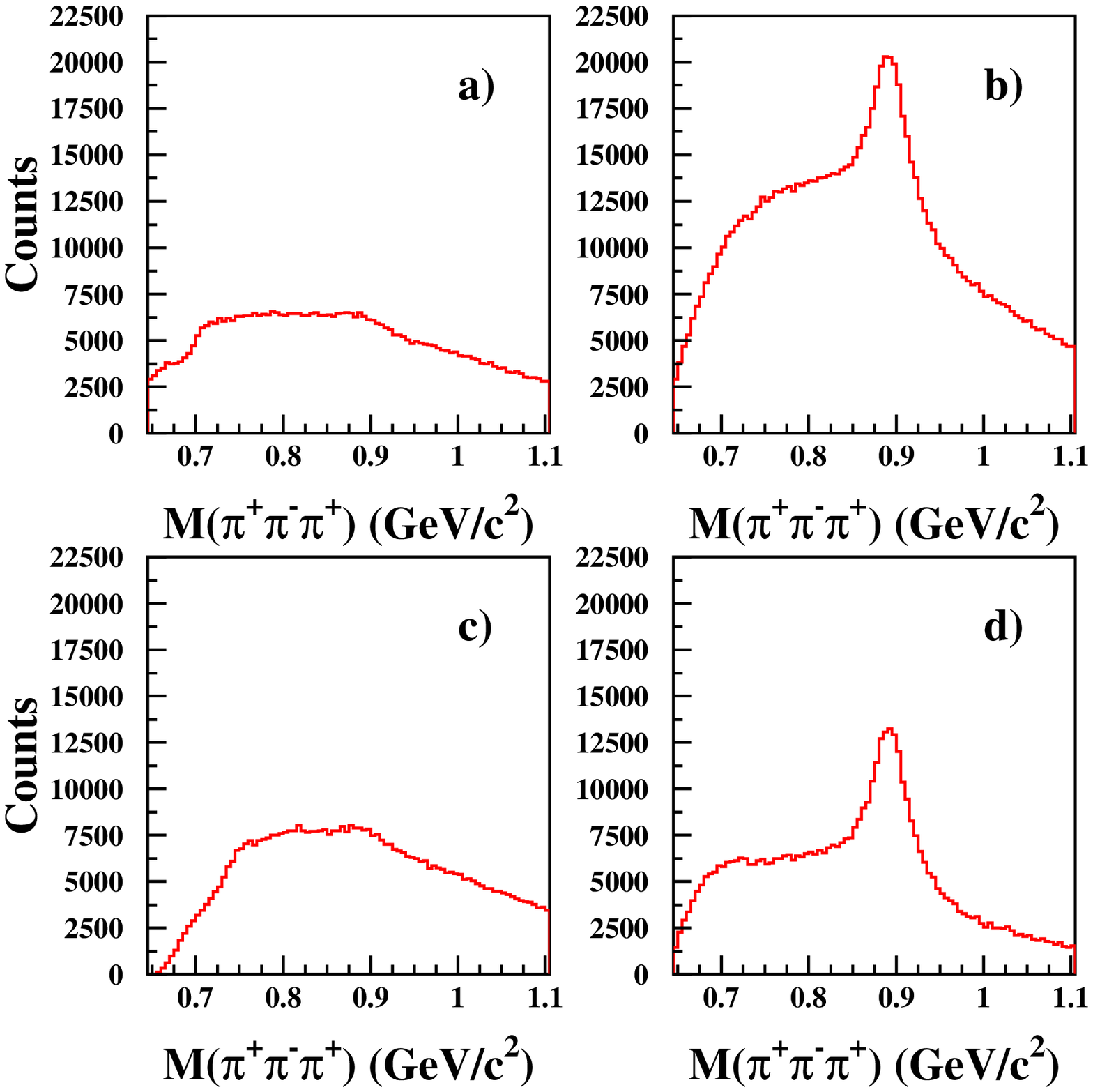,width=\columnwidth}
\epsfig{file=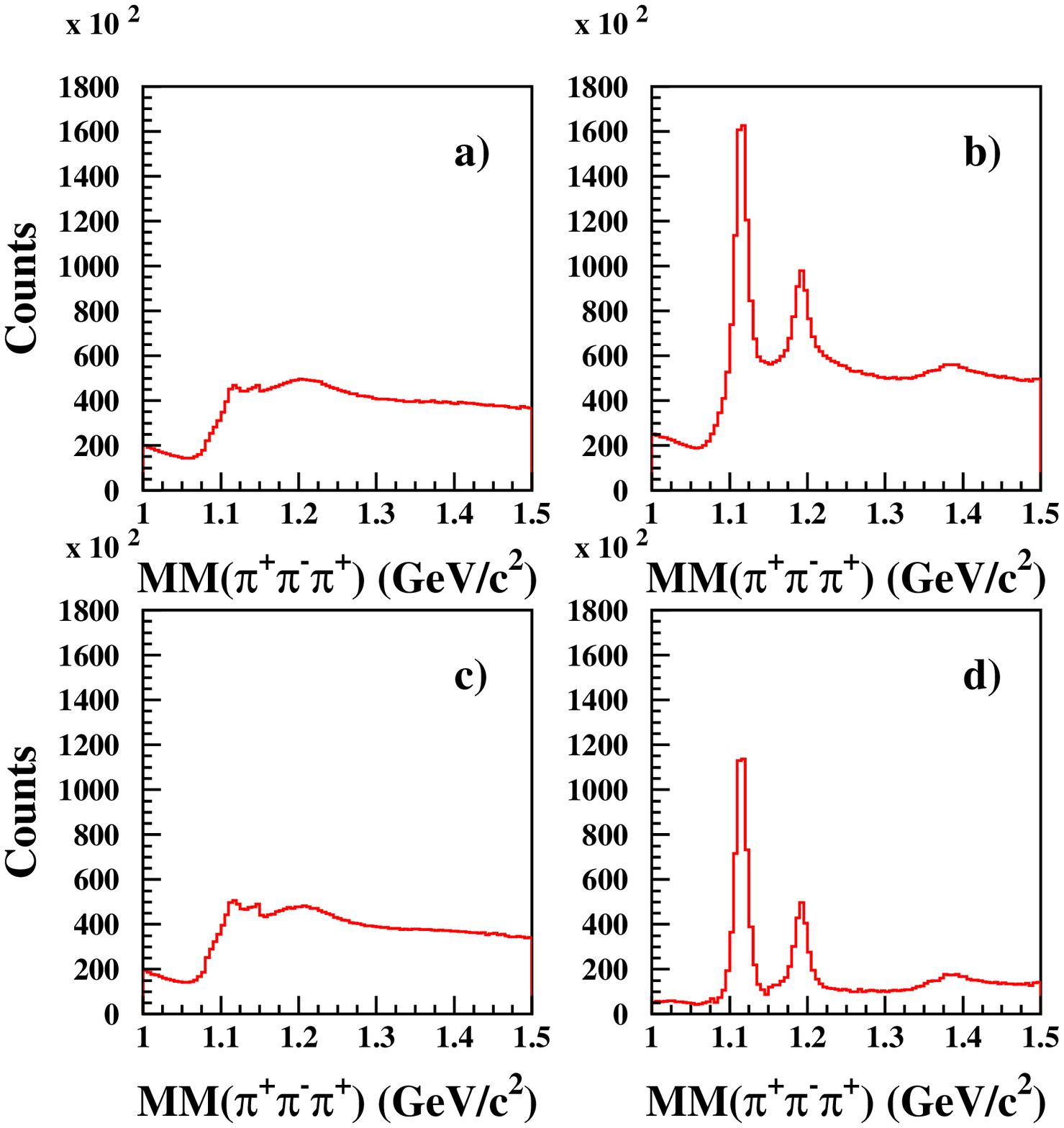,width=\columnwidth}
\caption{Three-pion invariant mass (left) and 
the three-pion missing mass (right). 
The four plots in each group correspond to: 
a) the left band, 
b) the middle band before the SSM, 
c) the right band and 
d) the middle band after the SSM. 
The peak in the three-pion mass is the $K^{*+}$ and the peaks 
in the three-pion missing mass are the $\Lambda$ and $\Sigma^{0}$. 
All plots are integrated over all incident photon energies.}
\label{sideband}
\end{figure*}

Fig.~\ref{sideband} shows the sideband subtraction applied to the 
reconstructed three-pion invariant mass and to the three-pion missing mass. 
The SSM reduces the background, giving cleaner signal peaks.

\subsection{Peak Fitting}

After applying the SSM to each $M(\pi^+\pi^+\pi^-)$ 
invariant mass plot, corresponding to different incident photon energy 
and different $K^{*+}$ production angle ranges, the $K^{*+}$ peak 
becomes clearer, but it is still not free of background. 
The main contribution to the background comes from the reaction channel  
$\gamma p \to K^{0} \Sigma^{*+}$(1385), which passed 
through  all the cuts. In addition, the 3-body phase space reaction 
$\gamma p \to K^{0} \pi^{+} \Lambda$  is also present, 
and will contribute to the background as well. 

In order to extract the correct $K^{*+}$ peak yield, 
instead of fitting  the  $K^{*+}$ peaks directly with a 
Breit-Wigner plus background functions, we applied a template 
fit. The precondition for this template fitting 
is that we assume there is negligible interference between the  
$K^{*+} \Lambda$ and $K^{0} \Sigma^{*+}$(1385) channels; 
in other words, we assume that the $K^{*+} \Lambda$ and 
$K^{0} \Sigma^{ *+}$(1385) add incoherently. If we remove all 
other sources of background, then the $K^{*+}$ mass plot should 
have background only from the $\Sigma^{*+}$(1385) peak. Similarly, 
the background in the $\Sigma^{*+}$(1385) plot comes only from 
events in the  $K^{*+}$ peak. 
Because the three-body $K^{0}$$\pi^{+}$$\Lambda$ channel is also a 
possible background, we assume it will add incoherently as well 
in both mass projections. 

To justify these assumptions, we explored the effect of various 
levels of interference between these two final states in the 
simulations.  The result is that the 
template fits correctly reproduced the generated events to within 
a 5\% uncertainty for assumptions of maximal constructive or 
destructive interference. 

\begin{figure}
\epsfig{file=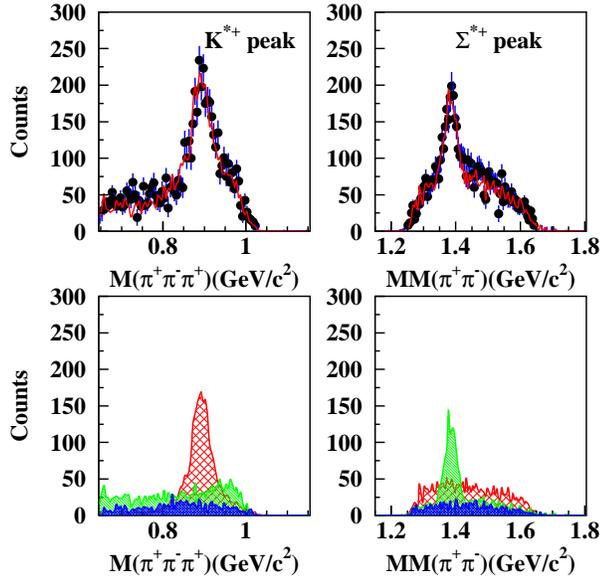,width=\columnwidth}
\caption{Example of the template fitting. Top: the solid dots 
are from the data, while the curve is from the fitting, which  
contains contributions from the $K^{*+} \Lambda$, $K^0 \Sigma^{*+}$ 
and $K^{0} \pi^{+} \Lambda$ channels, shown individually by 
the two plots at the bottom in large red diagonal cross, forward green diagonal  and small blue diagonal cross histograms, respectively.
}
\label{tpfit}
\end{figure}

Fig.~\ref{tpfit} shows an example of the template fitting, where the 
solid dots with error bars are from the data, while the curve is 
from the fit, which contains contributions from both the 
$K^{*+} \Lambda$, $K^0 \Sigma^{*+}$ and 
$K^{0} \pi^{+} \Lambda$ channels.
The $K^{*+}$ peak is seen in the left plots and the $\Sigma^{*+}$ 
peak is seen in the right plots.
The template shape for each contribution comes from the simulation 
for that channel, and the magnitude of each channel is a free parameter 
to optimize the fit, with the result for each component of the fit
shown in the bottom plots of Fig.~\ref{tpfit}. 
Both mass projections of Fig.~\ref{tpfit} are fit simultaneously 
to minimize the overall $\chi^2$.

\begin{figure}
\epsfig{file=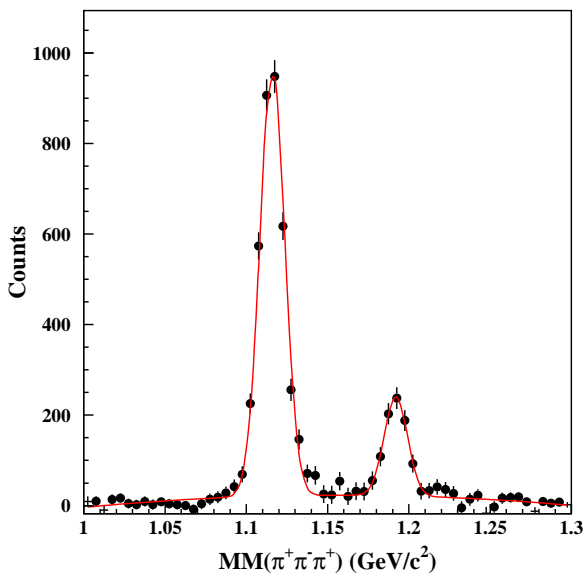,width=\columnwidth}
\caption{Example of two Gaussians plus a second order polynomial 
fit to the reconstructed $\Lambda$ and $\Sigma^{0}$ missing mass peaks.}
\label{fit_exp_kspsigma}
\end{figure}

For the $K^{*+} \Sigma^{0}$ reaction, the counts from the $\Sigma^{0}$ were extracted by using a Gaussian fit, then the yields were corrected bin by bin based on a Monte Carlo study of how much $K^{0} \Sigma^{\ast{}+}$(1385)  leakage there is to  $K^{*+} \Sigma^{0}$ reaction channel.  The correction was studied and found to be less than 0.1\%, which was included in our cross section calculation. Fig.~\ref{fit_exp_kspsigma} shows an example of the fitting. There are two peaks in the three-pion missing mass plot, one corresponding to the $\Lambda$ and the other to the $\Sigma^{0}$. The fitting function used two Gaussians plus a second order polynomial, for the $\Lambda$ peak, $\Sigma^{0}$ peak and background, respectively.

\subsection{Detector Acceptance}
\label{da}
A computational simulation package, the CLAS GEANT Simulation (GSIM),
was used for the Monte Carlo modeling of the detector acceptance. 
GSIM is based on the CERN GEANT simulation code with the CLAS 
detector geometry. 
Thirty million  $\gamma p \to K^{*+} \Lambda \ (\Sigma^{0})$ 
events were randomly generated, with all possible decay channels 
of the final state particles ($K^{*+}$, $\Lambda$, $\Sigma^{0}$\dots).
The Monte Carlo files were generated with a Bremsstrahlung photon 
energy distribution and a tunable angular distribution that best fit the $K^{\ast}$ data. 
The energy bin size was 0.1 GeV
and the total cross section was assumed constant across the bin. 
This assumption is reasonable based on the slowly-varying total 
cross sections shown below.
Because the simulations have a better resolution than the real CLAS 
data, the output from GSIM are put through a software program 
to smear the particle momentum,  timing, {\it etc.}  
to better match the real data.

An extensive study of the g11a trigger \cite{mike} showed a small 
inefficiency for the experimental trigger. To account for the trigger 
inefficiency, an empirical correction was mapped into the Monte Carlo.  
The trigger corrections applied here is the same as used for other 
CLAS analyses of this same dataset \cite{mike}.

The detector acceptance is calculated by:
\begin{equation}
\epsilon = \frac{D_{MC}}{G_{MC}}
\label{eq9}
\end{equation}
where $\epsilon$ represents the detector acceptance, 
$D_{MC}$ is the number of simulated events after processing and 
$G_{MC}$ is the number of generated events.  

The same software used for the experimental data was applied directly 
to the Monte Carlo data.  Simulated events are extracted by fitting 
each reconstructed $K^{*+}$ peak for a given photon energy and 
$K^{*+}$ production angle. 
In our analysis, a non-relativistic Breit-Wigner function 
\begin{equation}
{\vert}\mathcal{A}_{non-rBW}{\vert}^{2} = A\cdot\frac{\Gamma}{2\pi}\cdot\frac{1}{(E-E_{R})^{2}+\Gamma^{2}/4 },
\label{eq10}
\end{equation}
was used to extract the counts of the $K^{*+}$ peaks 
for $K^{*+} \Lambda$ channel. Here, $\Gamma$ is the full 
width at half maximum of the resonance peak, $E$ is the scattering 
energy and $E_{R}$ is the center of the resonance. 

As described in section \ref{cuts}, different methods were used for 
the $K^{*+} \Lambda$ and $K^{*+} \Sigma^0$ channels due to the 
presence of $\Lambda^*$ resonance contributions in the former.
For the $K^{*+} \Sigma^{0}$ channel, where there is no kinematic 
overlap from hyperon states, the counts under the $\Sigma^{0}$ peak 
were fitted directly using a Gaussian function.
Fitting the three-pion missing mass of the $\Sigma^{0}$ has less 
uncertainty than fitting the  $K^{*+}$ peak, since the $\Sigma^{0}$ 
peak is relatively narrow on top of a nearly flat background.
This method was used for both simulated and experimental data.

\section{Normalization and Cross Section Results}

The differential cross sections are calculated by the formula:
\begin{equation}
\frac{d\sigma}{d\cos\theta{}_{K^{\ast{}+}}^{CM}} = \frac{Y}
{N_{target} \cdot N_{gflux} \cdot \varepsilon
\cdot \Delta \cos \theta^{CM}_{K^{*+}} \cdot f_{lt}},
\label{eq14}
\end{equation}
where  $\frac{d\sigma}{d\cos \theta{}_{K^{\ast{}+}}^{CM}}$ is the differential cross section in the $K^{\ast{}+}$ angle center-of-mass (CM) frame, 
$Y$ is the experimental yield,  $N_{target}$ is the area density of protons 
in the target,  $N_{gflux}$ is the incident photon beam flux, 
$\varepsilon$ is the detector acceptance, 
$\Delta \cos \theta^{CM}_{K^{*+}}$ is the bin size in the 
$K^{*+}$ angle in the CM frame and 
$f_{lt}$ is the DAQ live time for the experiment.  

The detector acceptance $\varepsilon$ and experimental yields $Y$ for 
the $K^{*+} \Lambda$ and  $K^{*+} \Sigma^{0}$ reactions 
are described in the previous sections. 

For each incident photon beam energy range ($\Delta E = 0.1$ GeV), 
nine angular regions were measured, uniformly distributed between  
$-1.0 <  \cos \theta^{CM}_{K^{*+}} <  1.0$. 
Hence, $\Delta \cos \theta^{CM}_{K^{*+}}$ is $\frac{2}{9}$. 

The livetime $f_{lt}$ for the g11a experiment was carefully studied as 
a function of beam intensity, and found to be 0.82 $\pm$ 0.01 for this 
measurement \cite{battag}.   

In our analysis, photon flux was extracted in 
photon energy steps of 0.05 GeV.  In the final analysis, we used 
photon energy bins of 0.1 GeV, and the fluxes added appropriately.

The proton density $N_{target}$ is calculated using the formula:
\begin{equation}
N_{target} = \frac{\rho\cdot L\cdot N_{A}}{A},
\label{eq15}
\end{equation}
where $\rho$, $L$ and $A$ are the target density, target length and the 
atomic weight of hydrogen, respectively.  $N_{A}$ is Avogadro's number. 
For the g11a experiment, an unpolarized liquid hydrogen target was used. 
The target density $\rho$ was measured using:
\begin{equation}
\rho = a_{1}T^{2} + a_{2}P + a_{3}, 
\end{equation}
where $T$, $P$ are the target temperature and pressure (measured at the 
beginning of each CLAS run), while $a_{1}$, $a_{2}$, $a_{3}$ are the 
fitting parameters. The mean value of the target density $\rho$ for 
the g11a data was obtained by taking the average \cite{mike}:
\begin{equation}
\overline{\rho} = \frac{1}{N_{run}}\sum \rho_{r} = 0.07177 ~g/cm^{3}
\end{equation}
where $N_{run}$ is the number of runs.
Using the target length of 40 cm, this gives $N_{target}$.

\section{Results}

\begin{figure*}
\epsfig{file=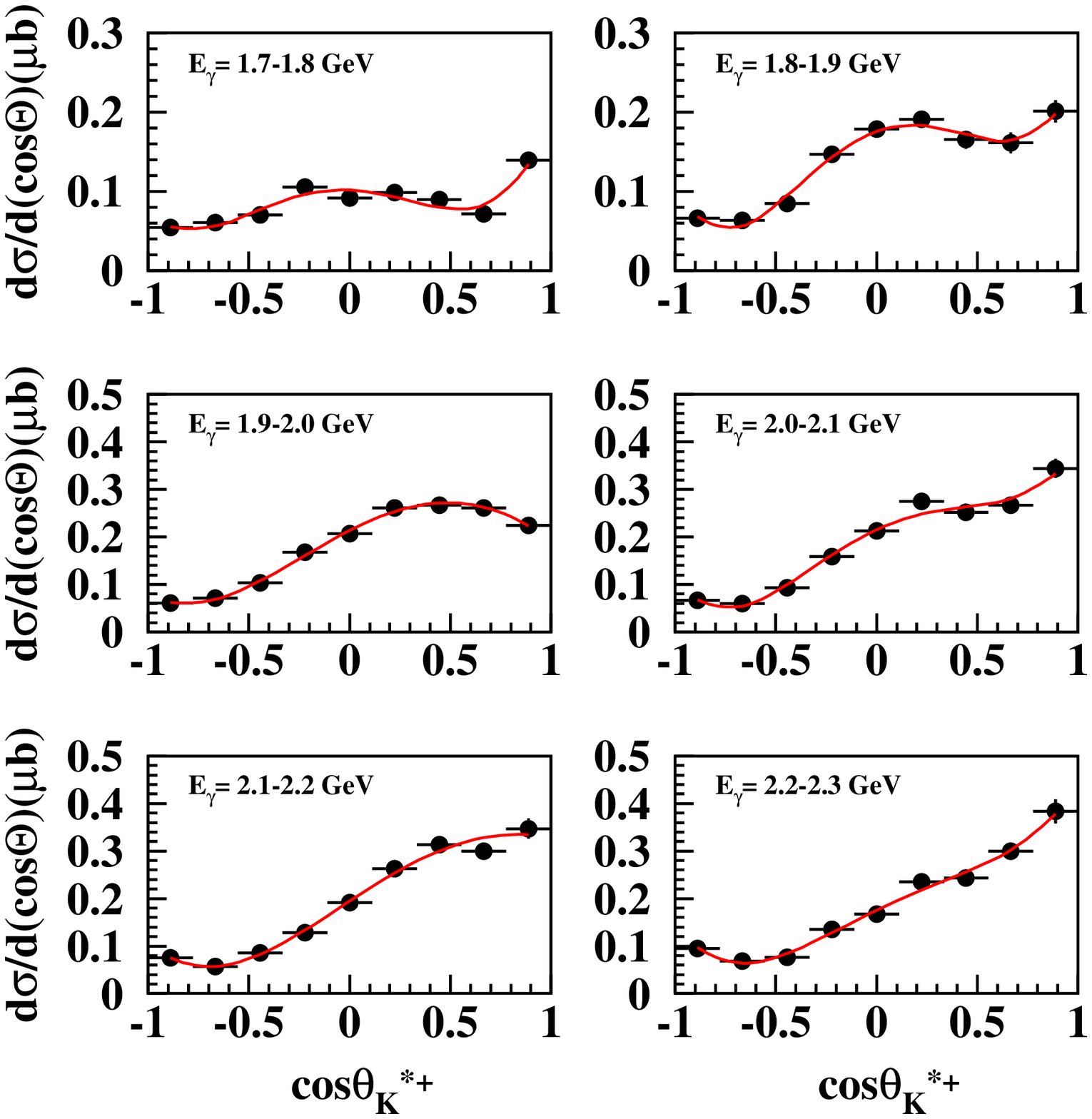,width=\columnwidth}
\epsfig{file=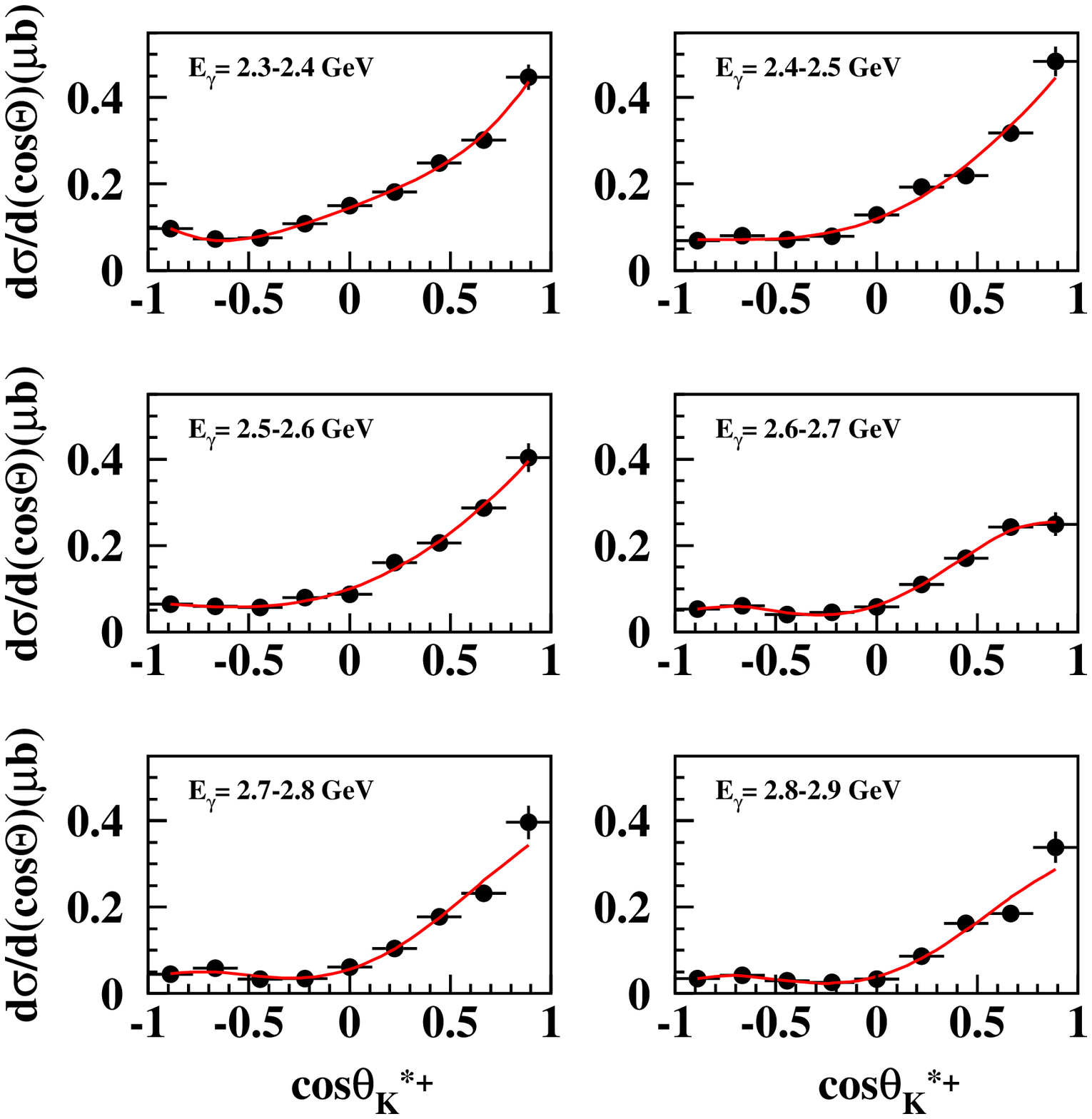,width=\columnwidth}
\epsfig{file=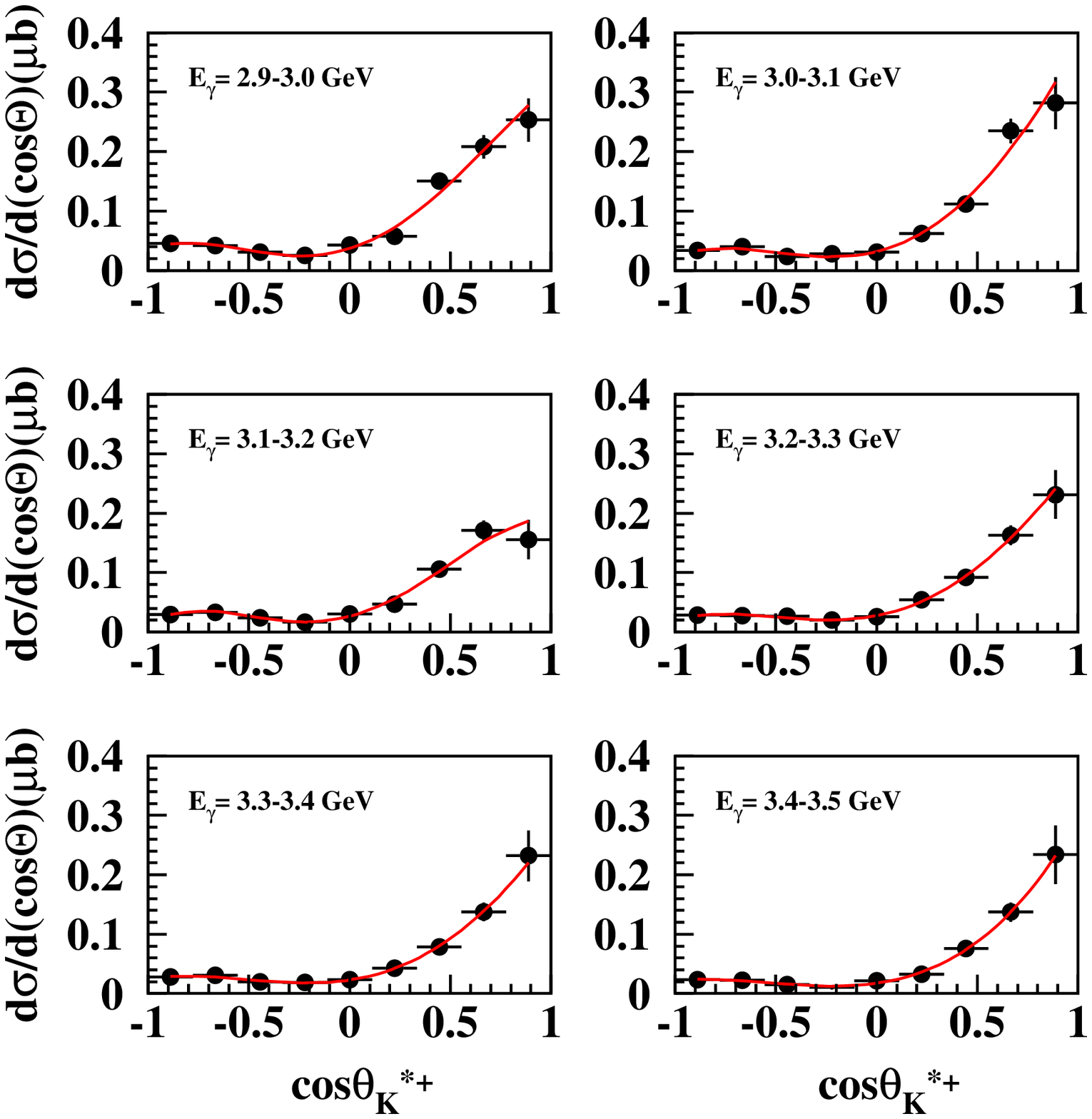,width=\columnwidth}
\epsfig{file=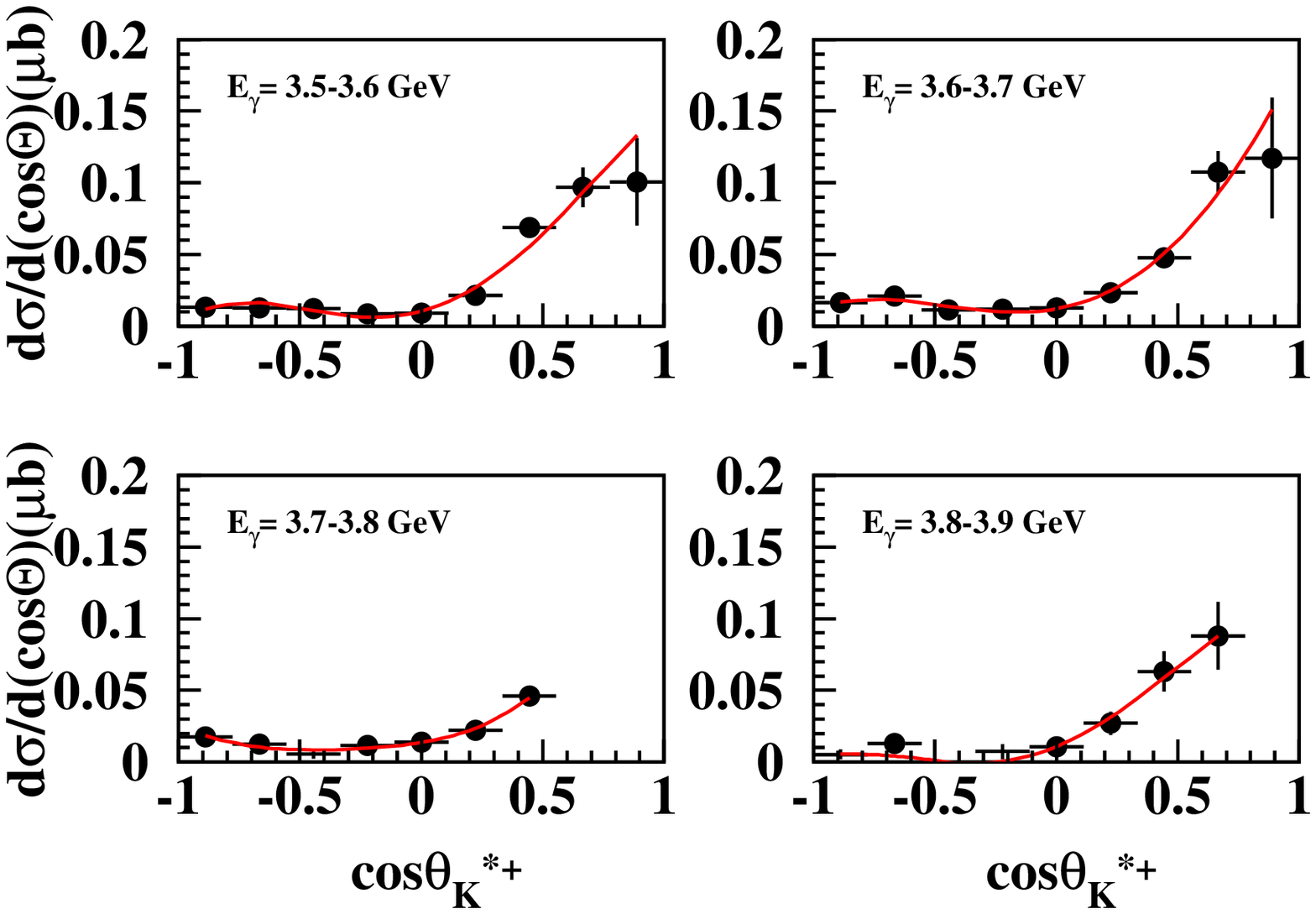,width=\columnwidth}
\hfill
\caption{Fitting the differential cross sections for $\gamma$$p$ $\to$ $K^{*+}$$\Lambda$  with 4th order Legendre polynomials. Incoming photon energies range from 1.7 to 3.9 GeV. }
\label{legfit_ksp1}
\end{figure*}

\begin{figure*}
\epsfig{file=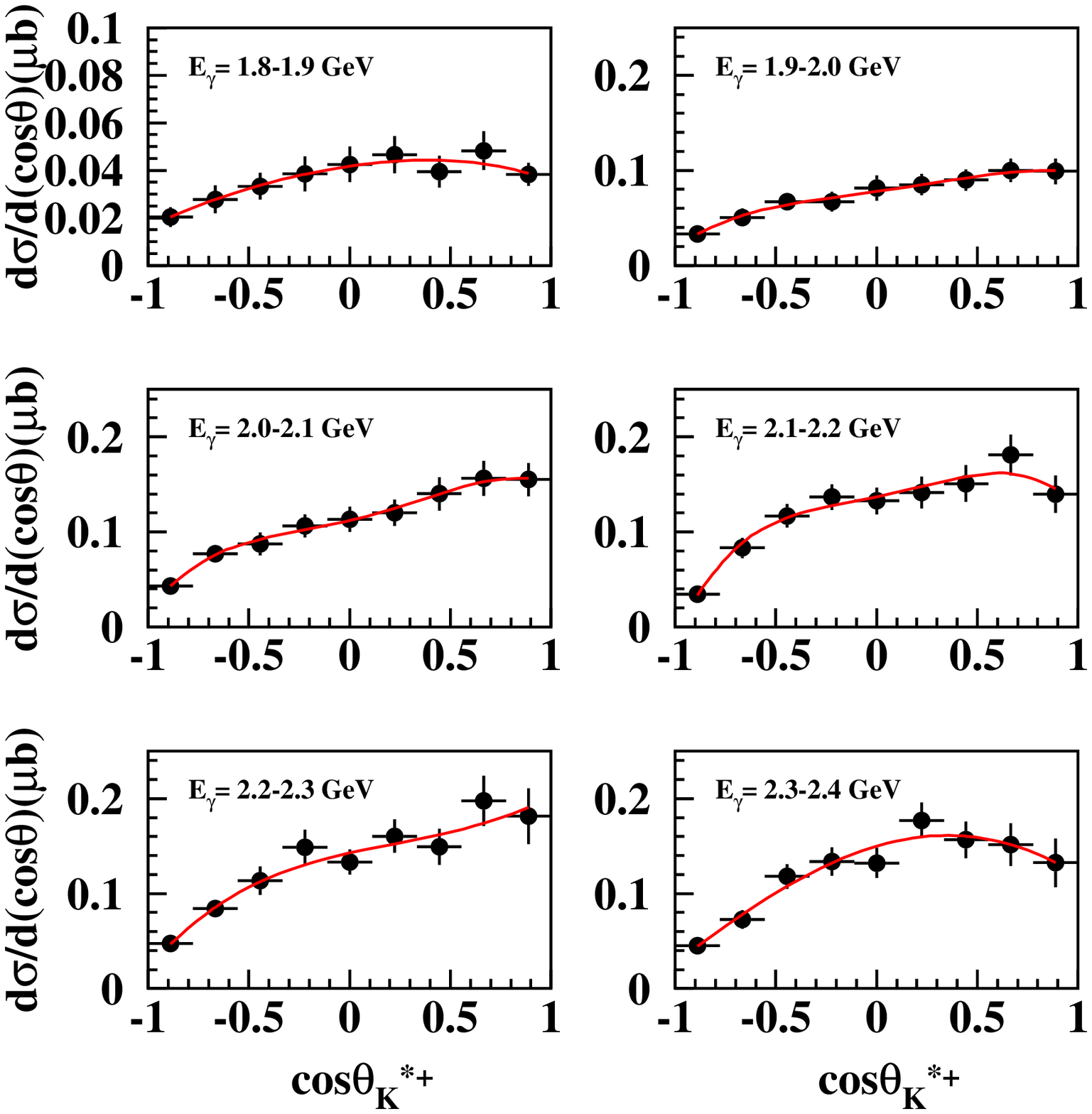,width=\columnwidth}
\epsfig{file=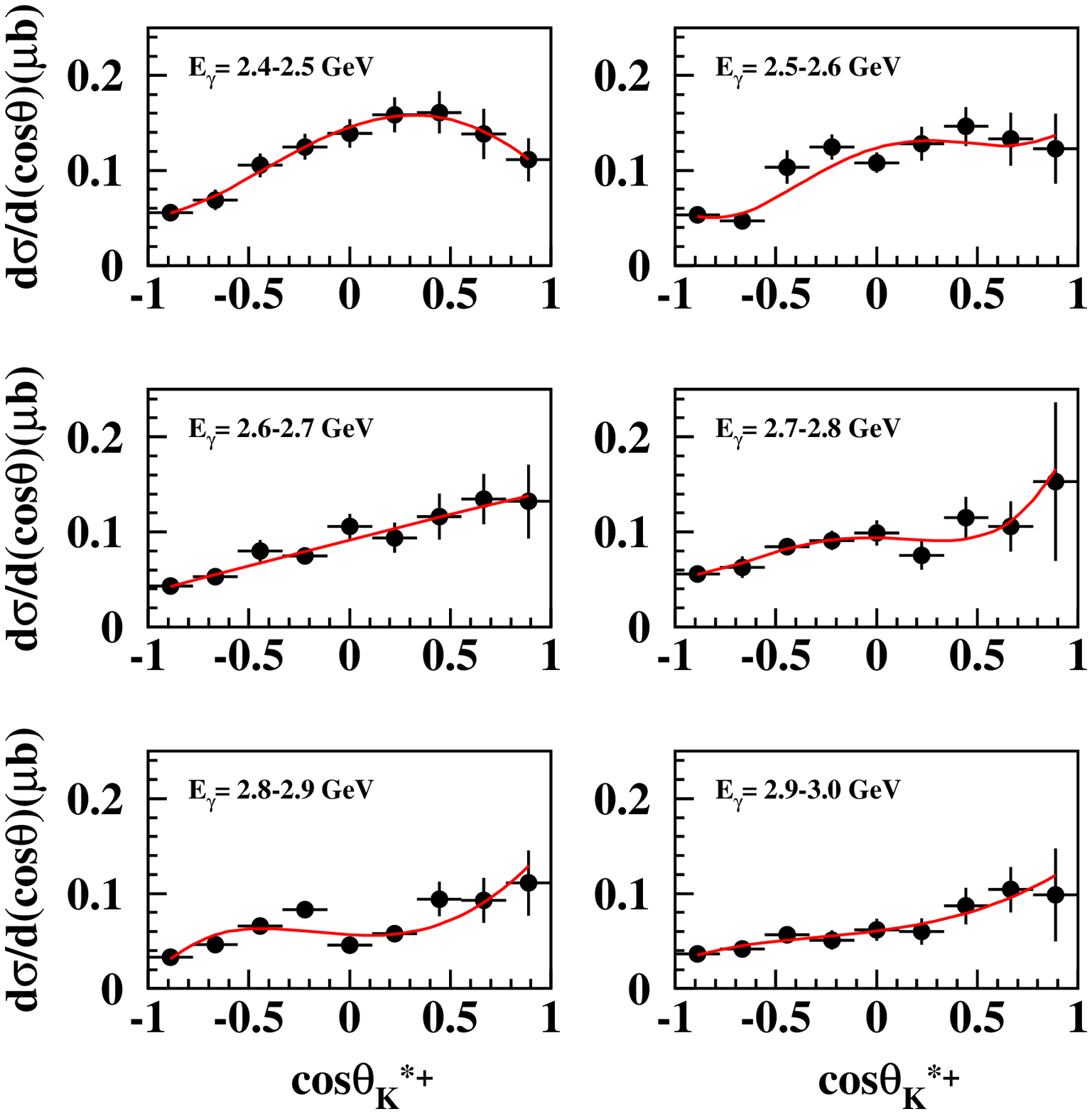,width=\columnwidth}
\epsfig{file=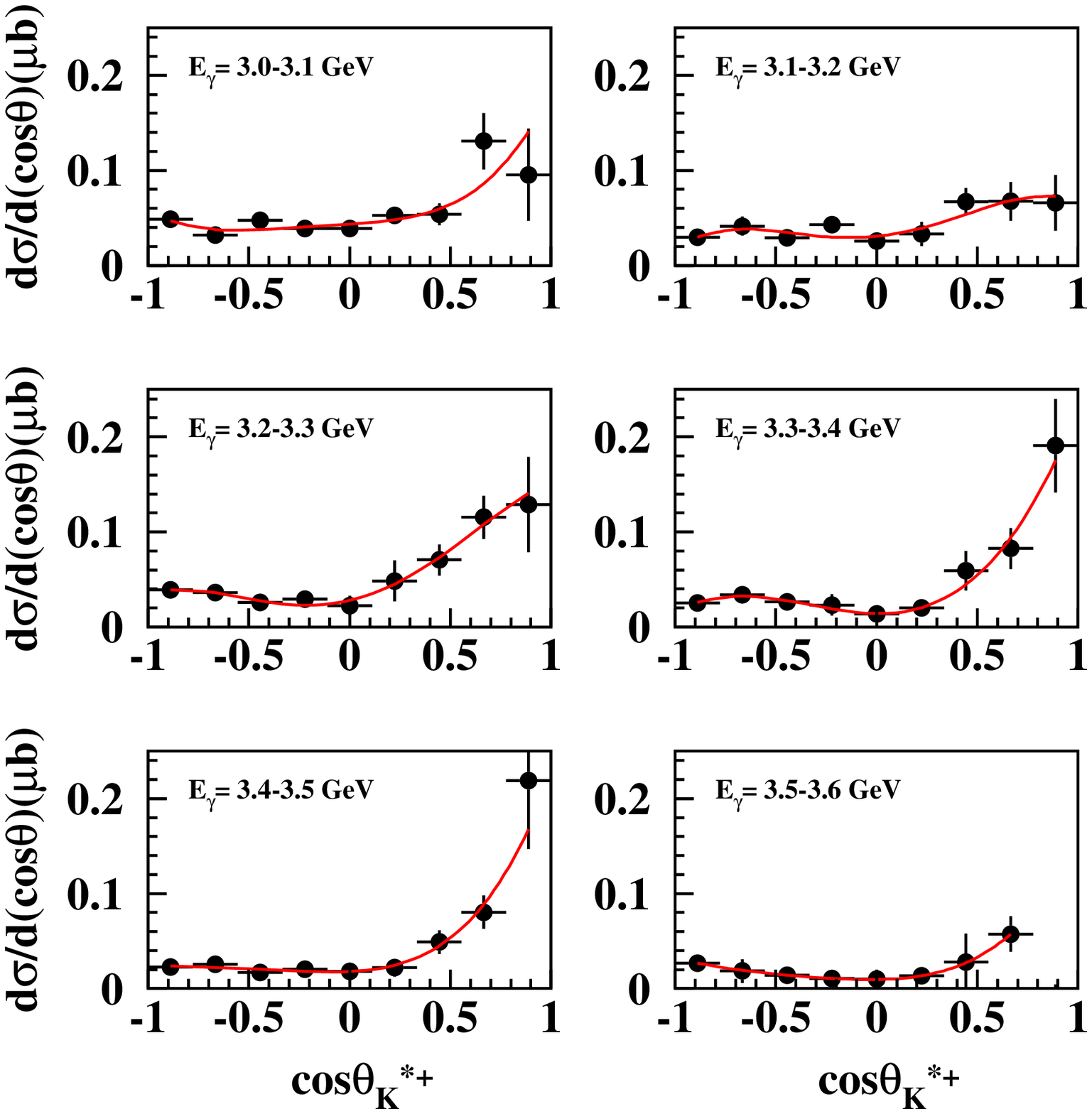,width=\columnwidth}
\epsfig{file=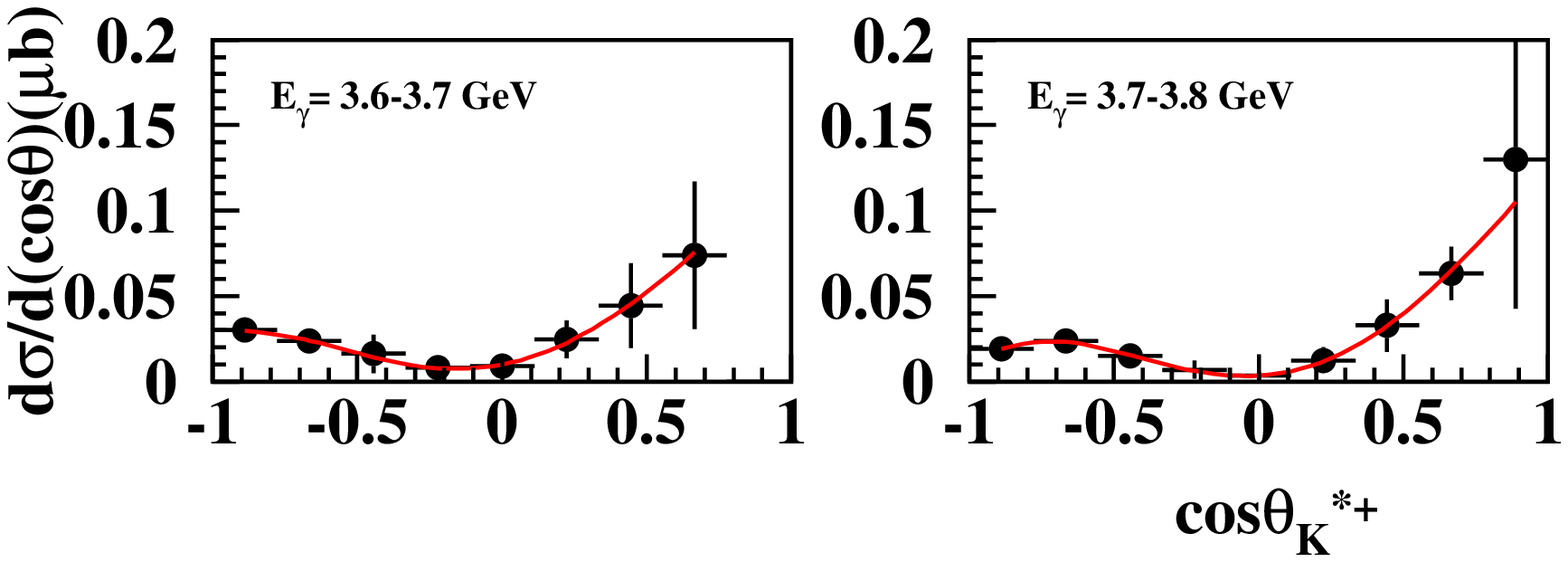,width=\columnwidth}
\caption{Fitting the differential cross section for  
$\gamma p \to K^{*+} \Sigma^{0}$ with 4th order Legendre polynomials. Incoming photon energies range from 1.8 to 3.8 GeV. }
\label{legfit_sig1}
\end{figure*}

Fig.~\ref{legfit_ksp1} 
shows the differential cross sections for the 
photoproduction reaction  $\gamma$$p$ $\to$ $K^{*+}$$\Lambda$, 
where there are 22 plots, for $E_{\gamma}$ bins ranging from 1.70 to 3.90 GeV. 
There are nine angular measurements in each plot, uniformly distributed 
in $\cos\theta^{CM}_{K^{*+}}$ between -1.0 and 1.0. 
In general, the $K^{*+}$$\Lambda$ differential cross sections 
shows dominantly a $t$-channel behavior, with an increase at forward-angles. 
Similarly, Fig.~\ref{legfit_sig1}  
shows the differential cross sections 
for  $\gamma$ $p$ $\to$ $K^{*+}$ $\Sigma^{0}$ photoproduction 
over the same photon energy range.
Comparison with theoretical calculations are given below in 
section \ref{theory}.

The differential cross sections can be decomposed into Legendre 
polynomials as \cite{bradford2}:

\begin{equation}
\frac{d\sigma}{d\cos\theta} = \frac{\sigma_{total}}{2} \lbrace 1 + 
\sum_{i=1}^{N} a_{i}p_{i}(x) \rbrace , 
\end{equation}
where $\sigma_{total}$ is the total cross section. 
By fitting the differential cross sections up to 4$^{th}$ order Legendre 
polynomials  
 \begin{equation}
f(x) = \sum_{i=0}^{4} a_{i}p_{i}(x),
\end{equation}
the total cross section was extracted by integrating  $f(x)$ over 
$\cos \theta$ from -1 to 1. Using the properties of the Legendre polynomials, 
after the integration,  only the $a_{0}$ term is left. 
Hence the total cross section is given by
$\sigma_{total} = 2 \cdot a_{0}$

\begin{figure*}
\epsfig{file=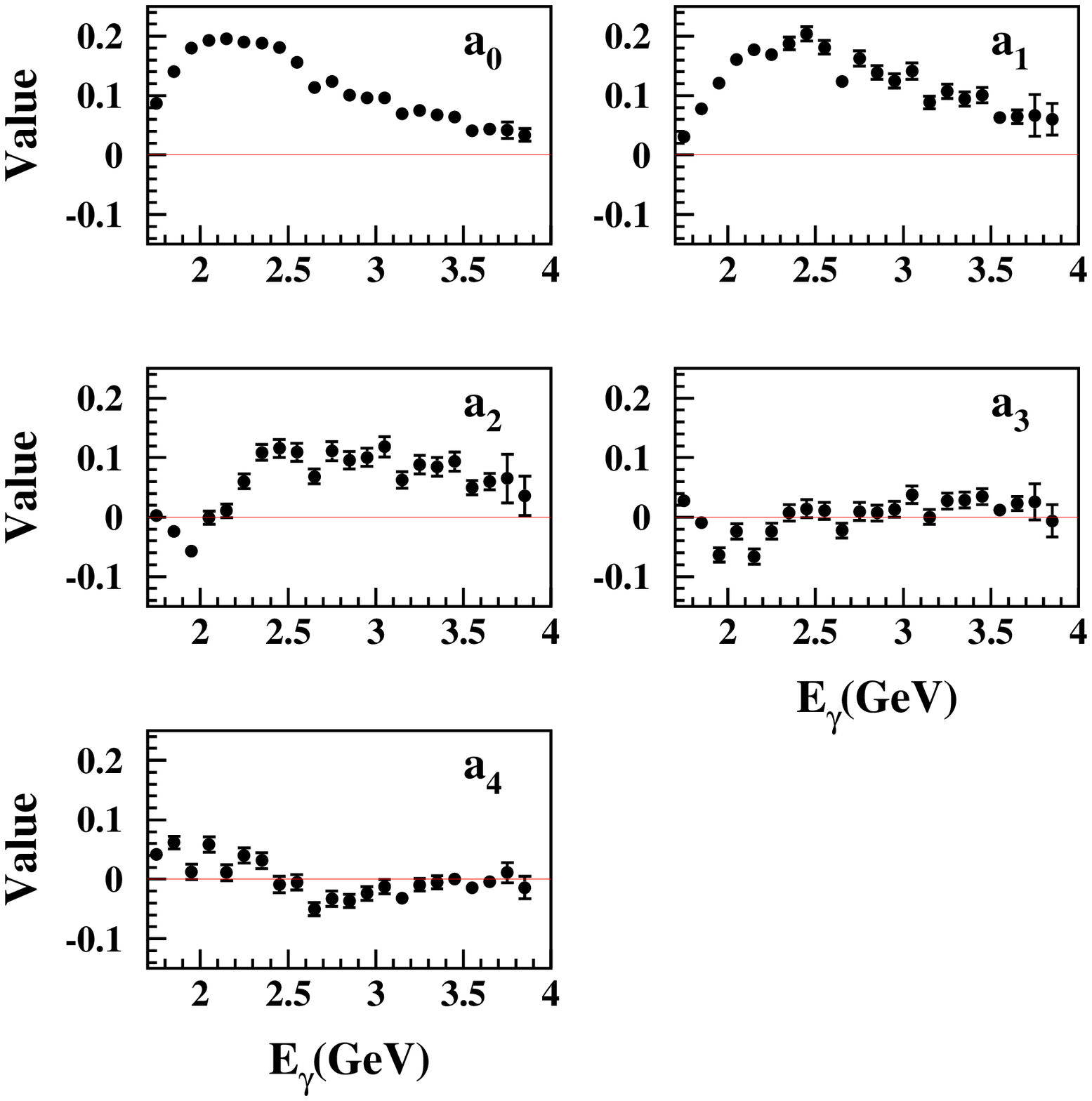,width=\columnwidth}
\epsfig{file=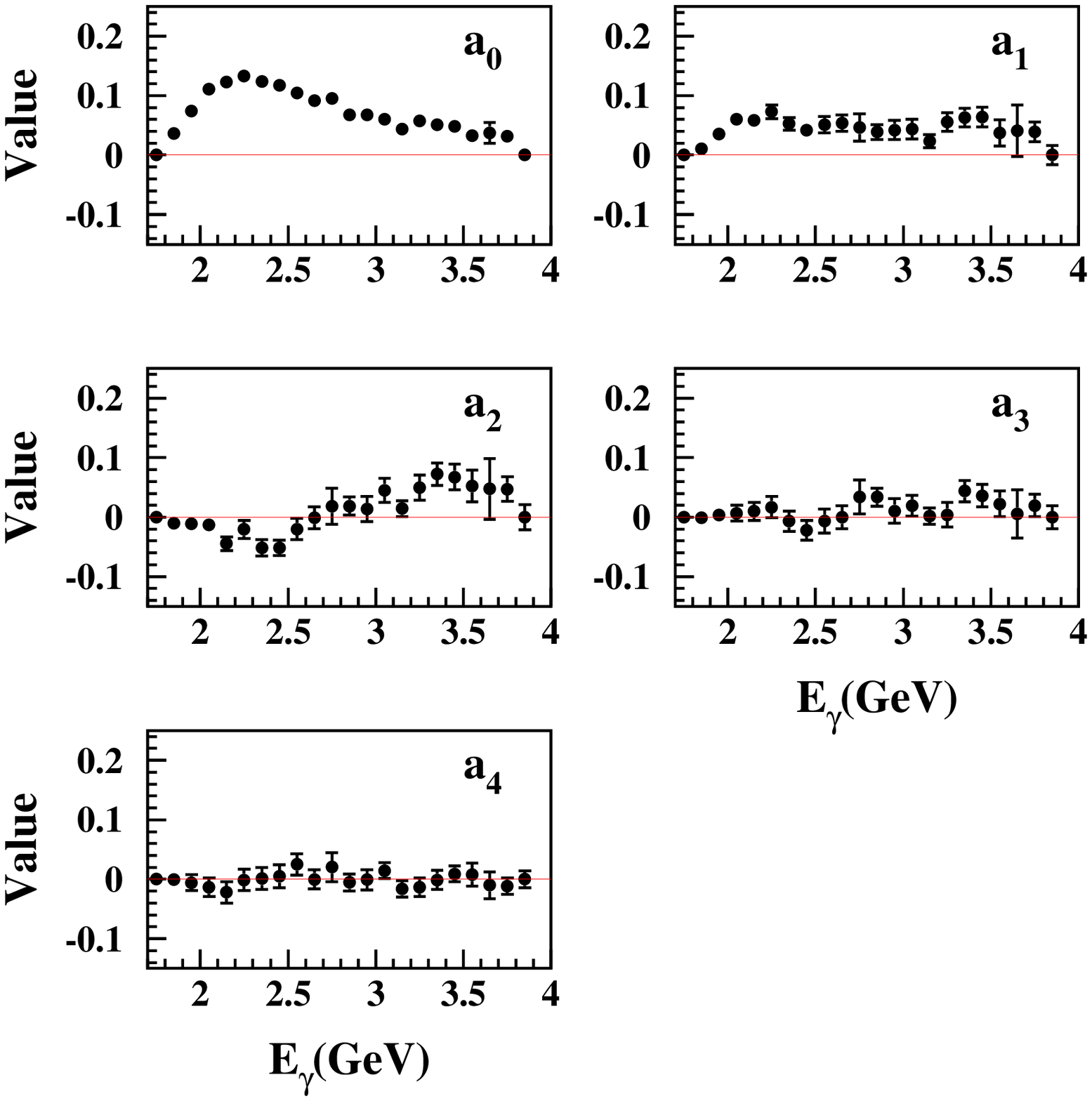,width=\columnwidth}
\caption{Legendre polynomial fitting parameters up to 4$^{th}$ order 
plotted versus incident photon energy $E_\gamma$ for $\gamma$ $p$ $\to$ $K^{*+} \Lambda$ (left) and  $\gamma$ $p$ $\to$  $K^{*+} \Sigma^{0}$ (right).}
\label{legen_parameter1}
\end{figure*}

\begin{figure*}
\epsfig{file=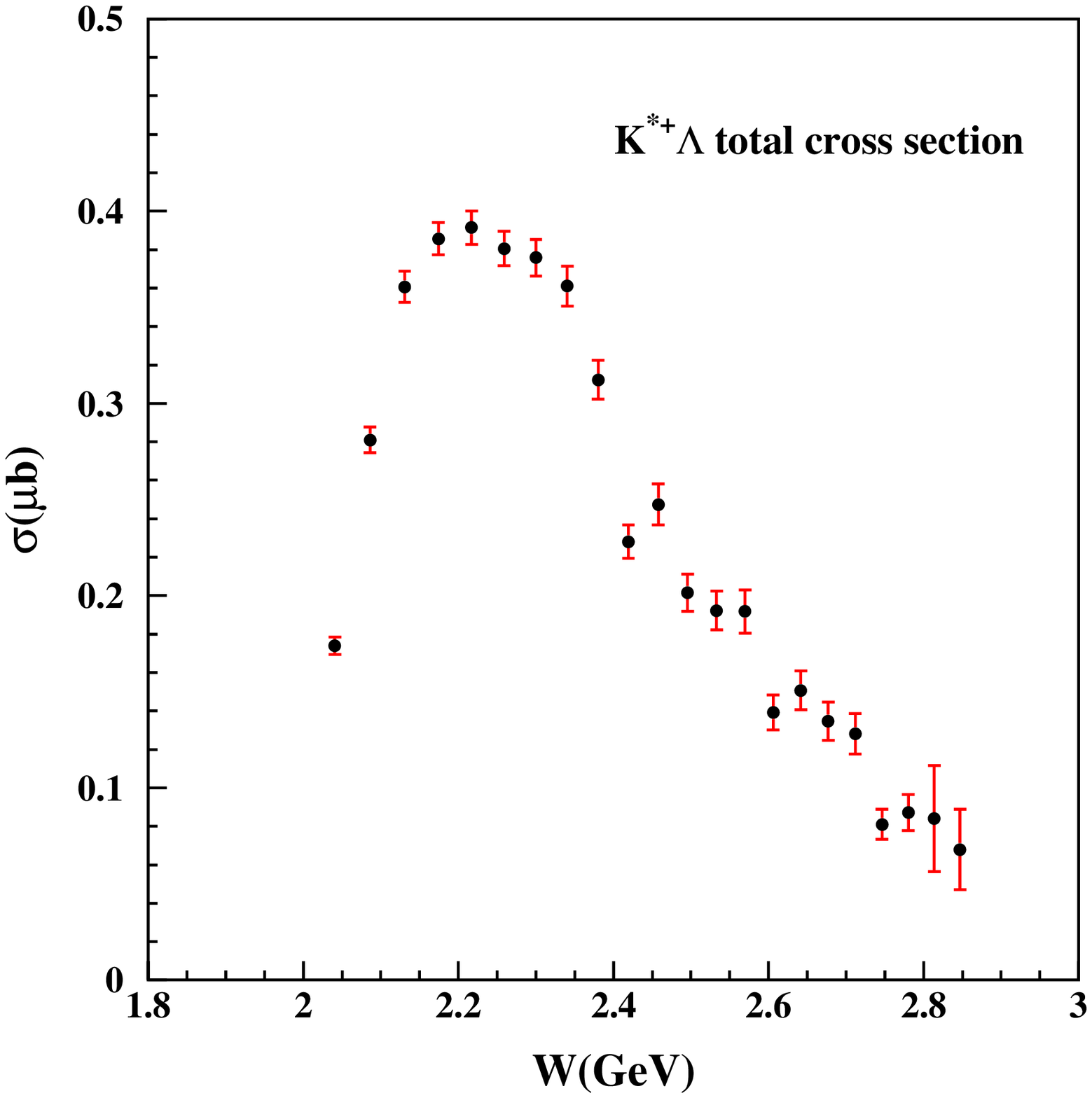,width=\columnwidth}
\epsfig{file=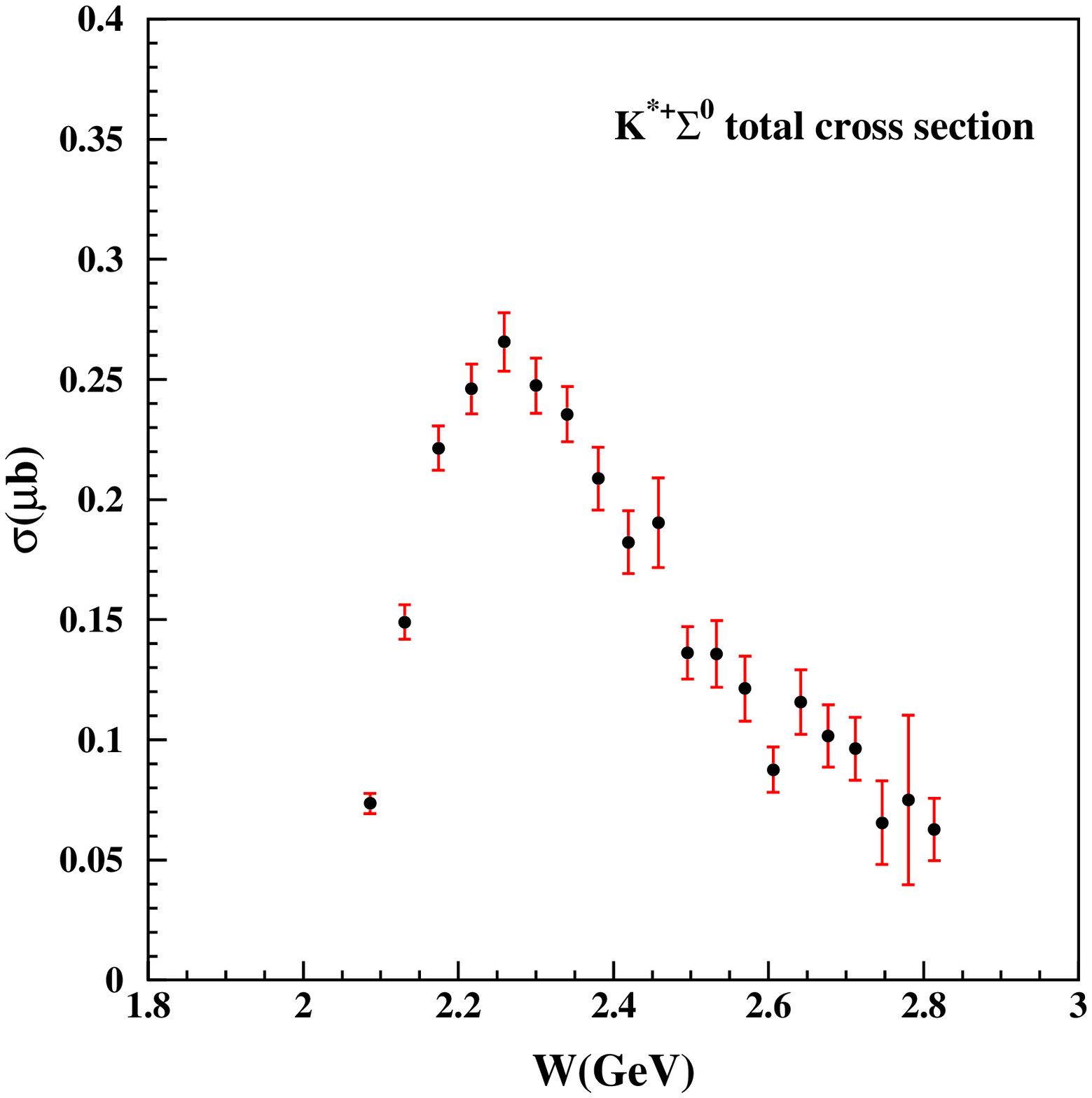,width=\columnwidth}
\caption{Total cross sections of the reaction $\gamma p \to K^{*+} \Lambda$ (left) and $\gamma p \to K^{*+}\Sigma^0$ (right).}
\label{cross_lam}
\end{figure*}

Fig.~\ref{legfit_ksp1} shows the fitting for the 
$\gamma$$p$ $\to$ $K^{*+}$$\Lambda$ channel, and 
Fig.~\ref{legfit_sig1} shows the fits for the  
$K^{*+}$$\Sigma^{0}$ final state. 
The fitting parameters $a_{0}$ through $a_{4}$ for each channel are plotted 
versus the incident photon energy $E_{\gamma}$ in 
Fig.~\ref{legen_parameter1}. 
The extracted total cross sections  are shown in 
Fig.~\ref{cross_lam} for 
the $K^{*+} \Lambda$ and  $K^{*+} \Sigma^{0}$ final states 
along with some theoretical curves explained below.  
The error bars show only the statistical uncertainty.

\subsection{Systematic Uncertainties} 

Systematic uncertainties come from several sources: the applied cut parameters, the choice of fitting functions, the Monte Carlo used for the 
detector acceptance and so on. 

Systematic uncertainties were estimated for each cut by varying the 
cut intervals and then recalculating the differential cross sections. 
The changes to cut parameters were applied to both the experimental 
data and the simulated output. The relative difference between the 
new cross sections and the original cross sections was calculated 
bin by bin using:
\begin{equation}
\delta\sigma = \frac{\sigma_{new} - \sigma_{old}}{\sigma_{old}}
\end{equation}
and then the resulting $\delta \sigma$ values were histogrammed.
This histogram was fitted with a Gaussian function, and the width from 
the Gaussian fit was taken as the systematic uncertainty for each variation. 
The cut intervals were varied to both larger and smaller values, and we chose 
the larger of the systematic uncertainties calculated from each variation. 

Similar estimation were done for the detector acceptance, by varying the 
inputs to the Monte Carlo. Also, different fitting functions and background 
shapes were used to determine the systematic uncertainties associated with 
the peak yields. The total systematical uncertainty is then given by 
\begin{equation}
\delta_{total} = \sqrt{\delta_{a}^{2} + \delta_{b}^{2} + \delta_{c}^{2} 
+ \dots} . 
\end{equation}
which assumes no correlated uncertainties.

The total systematic uncertainty from all sources, added in quadrature, 
is shown in Table \ref{sysana_all}, where the other sources include the 
target length, density and so on. For the $K^{*+} \Lambda$ final state 
the overall systematic uncertainty is 14\% and for  
$K^{*+} \Sigma^{0}$ the systematic uncertainty is 12\%.
   
\begin{table}
\caption{Summary of systematic uncertainties.}
\begin{center}
\begin{tabular}{lcc}
\hline
 &  $K^{*+} \Lambda$ channel    &   $K^{*+}\Sigma^{0}$ channel \\
\hline
Event Selections 	& 2.9\% & 4.5\% \\
\hline
Peak Fitting 		& 7.4\% & 5.8\% \\
\hline
Detector Acceptance  	& 9.2\% & 5.7\% \\
\hline
Beam Flux	  	& 7.0\% & 7.0\% \\
\hline
Other Sources 		& 2.5\% & 2.5\% \\
\hline
Total  			& 14\%  &  12\% \\
\hline
\hline
\end{tabular}
\end{center}
\label{sysana_all}
\end{table}

\subsection{Theoretical Calculations }
\label{theory}

The models that are currently available for $K^{*}$ photoproduction are 
based on effective Lagrangians, which fall into two groups: isobar models  
and Reggeized meson exchange models. 
Isobar models evaluate tree-level Feynman diagrams, which include 
resonant and nonresonant exchanges of baryons and mesons. 
The reggeized models, on the other hand, emphasize the $t$-channel 
meson exchange, which is expected to dominate the reaction at energies 
above the resonance region. The standard propagators in the Lagrangian 
are replaced by Regge propagators, which take into account an entire 
family of exchanged particles with the same quantum numbers instead of 
just one meson exchange. In this section, the $K^{*+} \Lambda$ cross 
section results will be compared with calculations from 
these two theoretical models.

One model we use is by Oh and Kim (O-K Model) \cite{oh2}, which is an isobar model.
This model starts with Born terms, which include $t$-channel (with $K$, 
$K^{*}$ and $\kappa$ exchanges), $s$-channel ground state nucleon 
exchanges and $u$-channel $\Lambda$, $\Sigma$ and $\Sigma^{*}$ exchanges. 
Additional $s$-channel nucleon resonance exchanges were added to the model 
using the known resonances from the PDG in Ref. \cite{KNOK}, referred to 
here as the K-N-O-K model. 
One attractive point of these models is the inclusion of diagrams with a light 
$\kappa$ meson exchange in the $t$-channel. As mentioned in the introduction, 
the $\kappa$ meson has not yet been firmly established, and these models 
allows us to study the effect of possible $\kappa$ exchange.   

The other model shown here is the Ozaki, Nagahiro and Hosaka (O-N-H) Model 
\cite{ONH}, which is a reggeized model. 
This model takes into account all possible hadron exchanges with the 
same quantum numbers (except for the spin). The coupling constants and $\kappa$  exchange parameters are the same as those used in O-K Model \cite{oh2}. 

\begin{figure*}
\epsfig{file=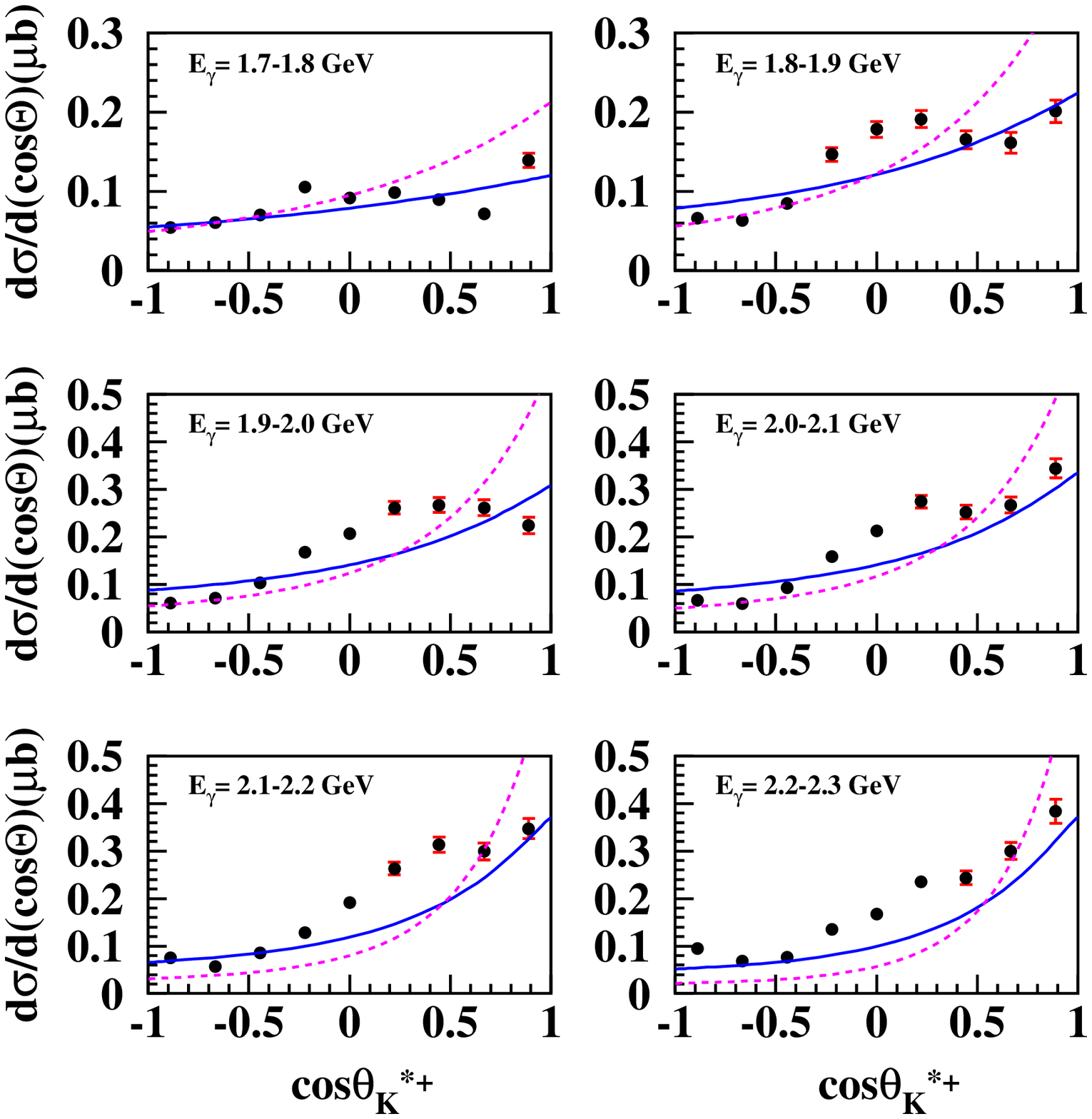,width=\columnwidth}
\epsfig{file=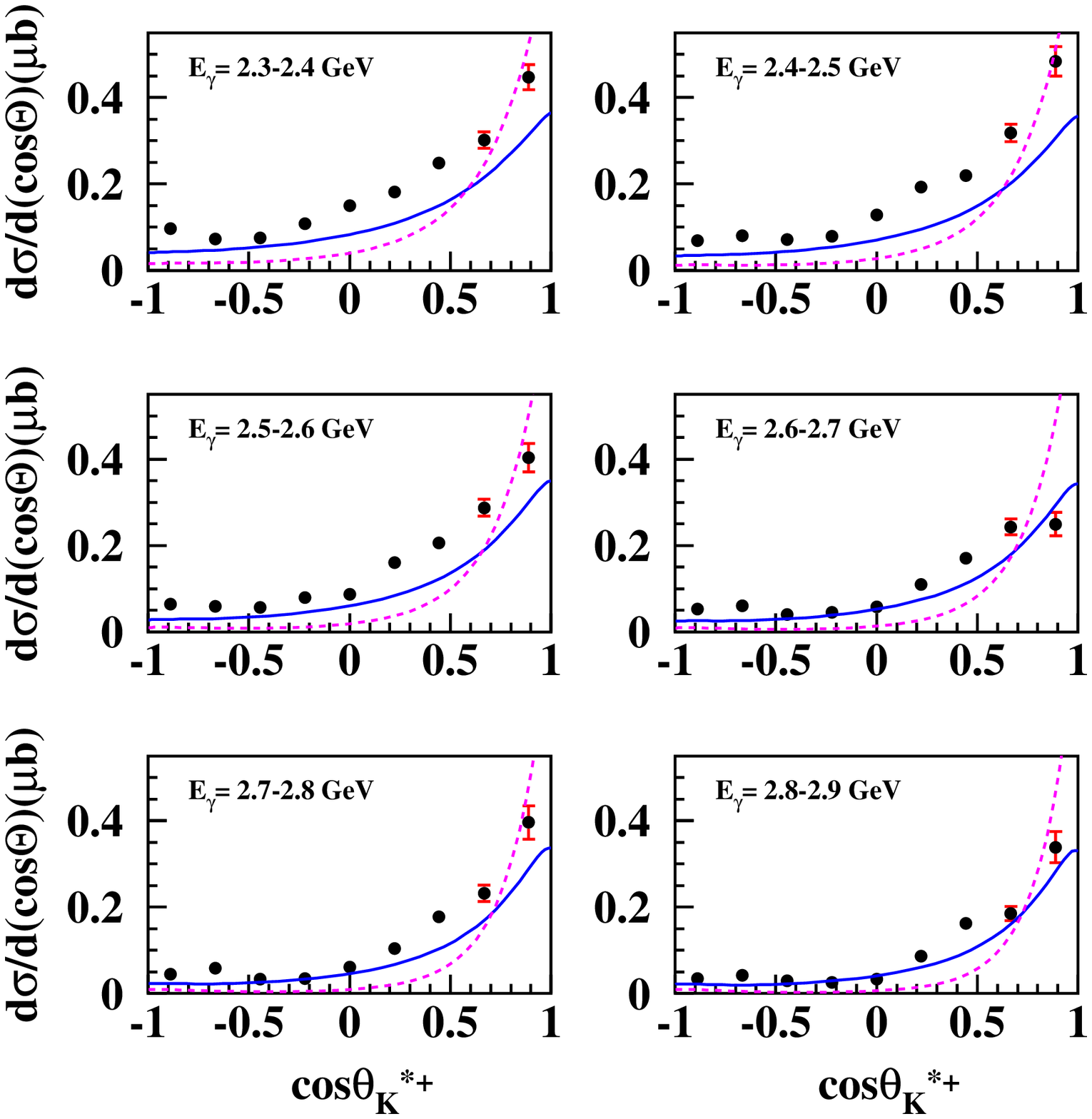,width=\columnwidth}
\epsfig{file=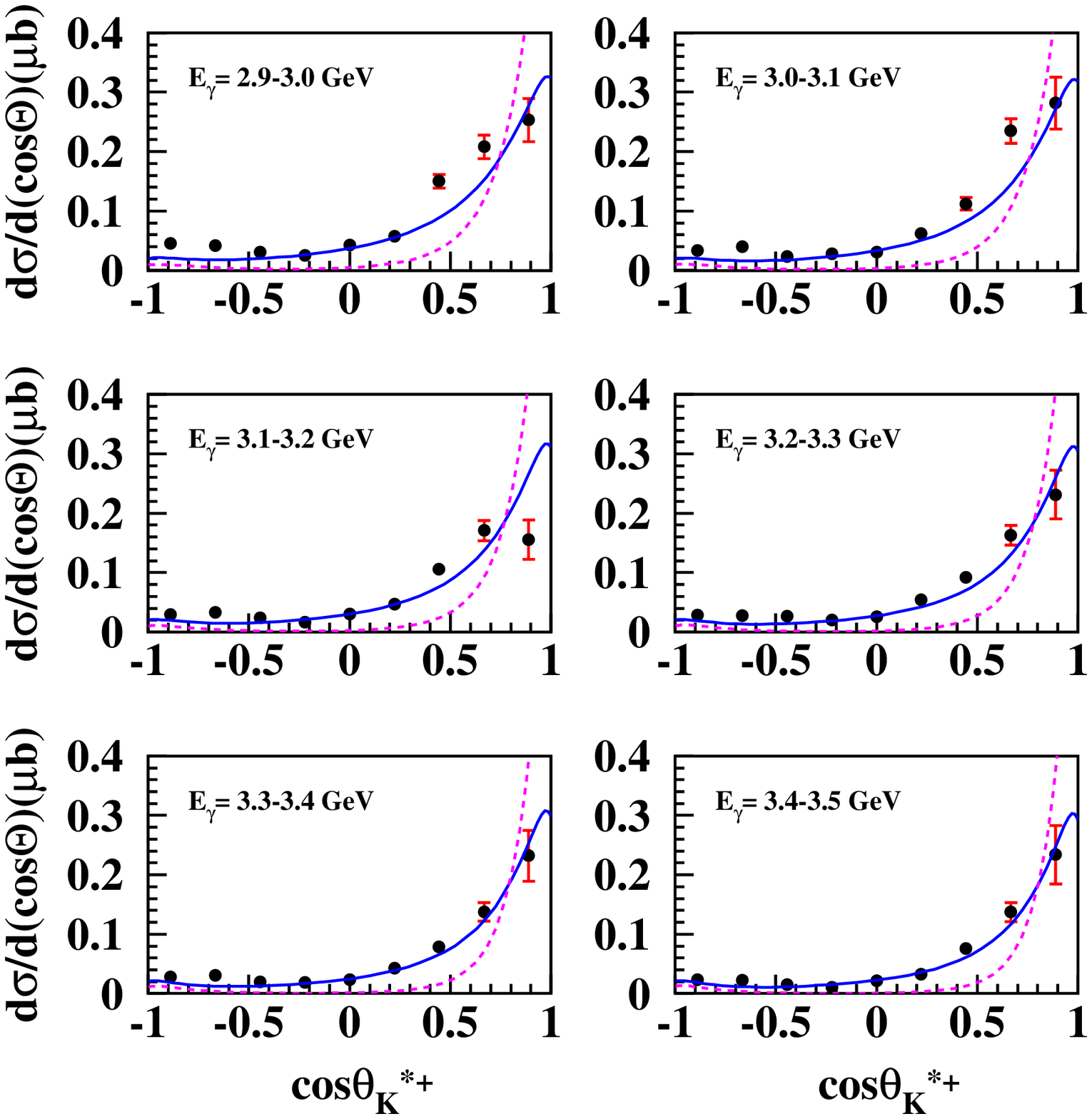,width=\columnwidth}
\epsfig{file=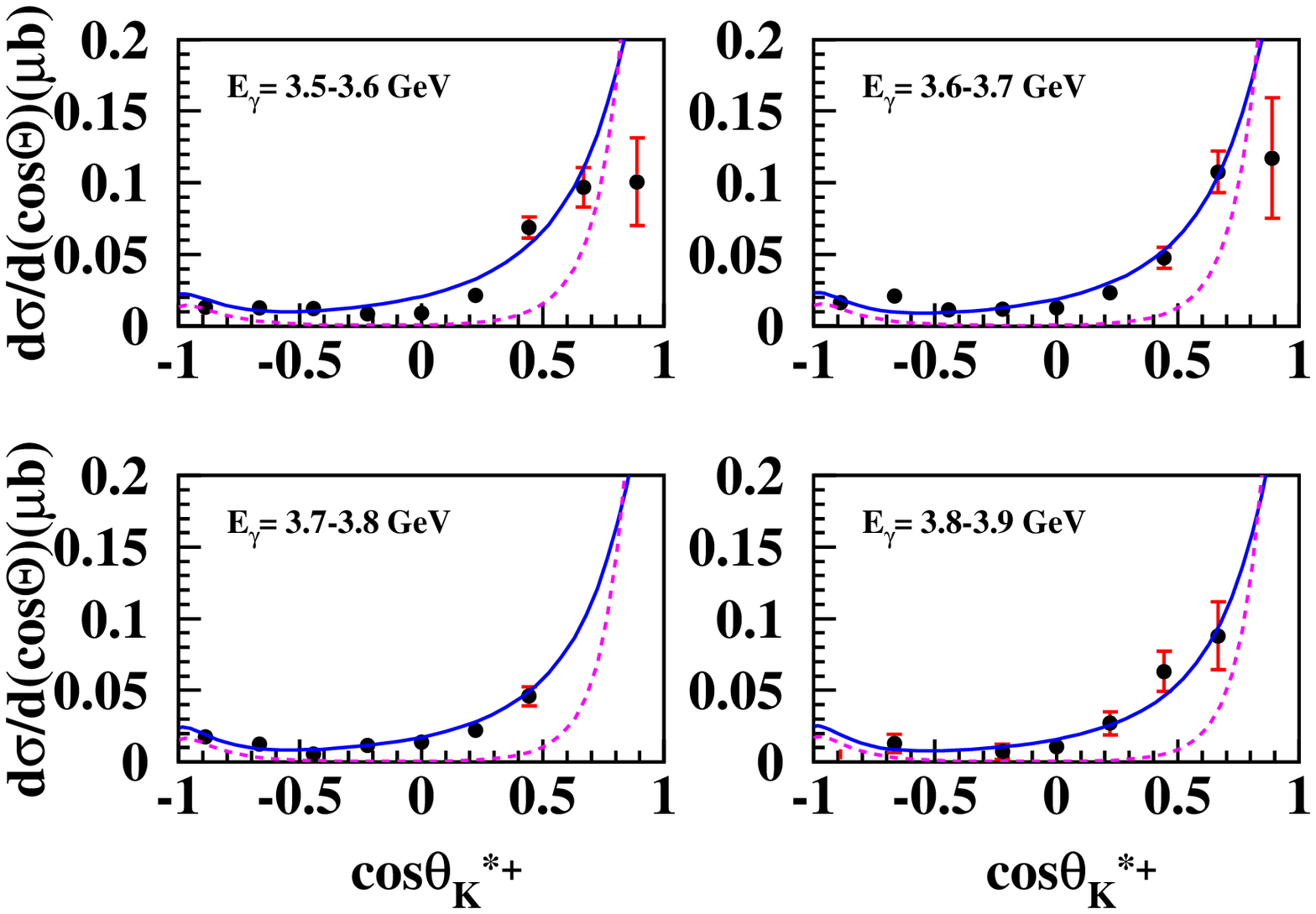,width=\columnwidth}
\caption{Differential cross sections of $\gamma p \to K^{*+} \Lambda$ 
plotted for incident photon energies from 1.7 to 3.9 GeV. 
The solid (blue) curves represent the theoretical calculations from the 
K-N-O-K Model \protect \cite{KNOK}
and the dashed (magenta) curves represents the calculations from 
O-N-H Model with resonance terms \protect \cite{ONH}. }
\label{difcross_ksp_theory1}
\end{figure*}

\begin{figure}
\epsfig{file=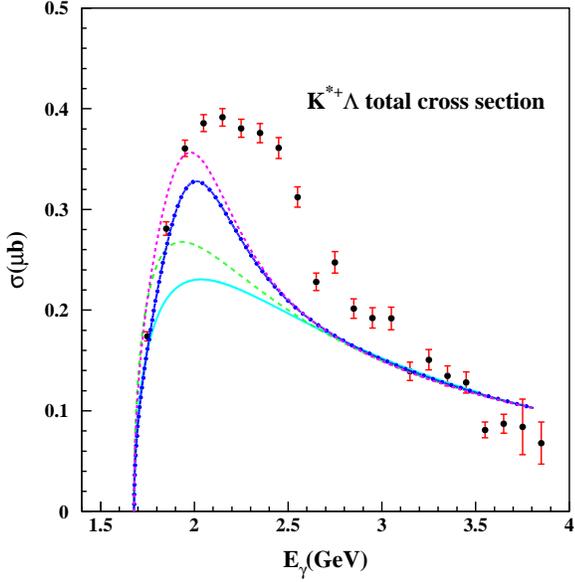,width=\columnwidth}
\caption{Total cross sections of the reaction $\gamma p \to K^{*+} \Lambda$.
 The solid (cyan) and  dash-dotted (blue) curves represent the theoretical 
calculations from the O-K and K-N-O-K models, respectively.  
The dotted (magenta) and dashed (green) curves represents the 
O-N-H model with and without resonance terms. }
\label{cross_lam_theory}
\end{figure}

Fig.~\ref{difcross_ksp_theory1}    
shows those calculations compared with our 
differential cross sections, where the solid curves 
represent the theoretical calculations from the K-N-O-K Model
and the dashed curves represent the O-N-H Model.
The corresponding curves are shown in 
Fig.~\ref{cross_lam_theory} for the total cross sections, where 
the curves are explaind in the figure caption.
The O-K model includes $s$-channel diagrams with most well-established 
nucleon resonances below 2 GeV \cite{pdg}, whereas the K-N-O-K model 
includes two additional $s$-channel resonances up to 2.2 GeV.  
Interpretation of these results are discussed in section \ref{conclude}.

\begin{figure}
\epsfig{file=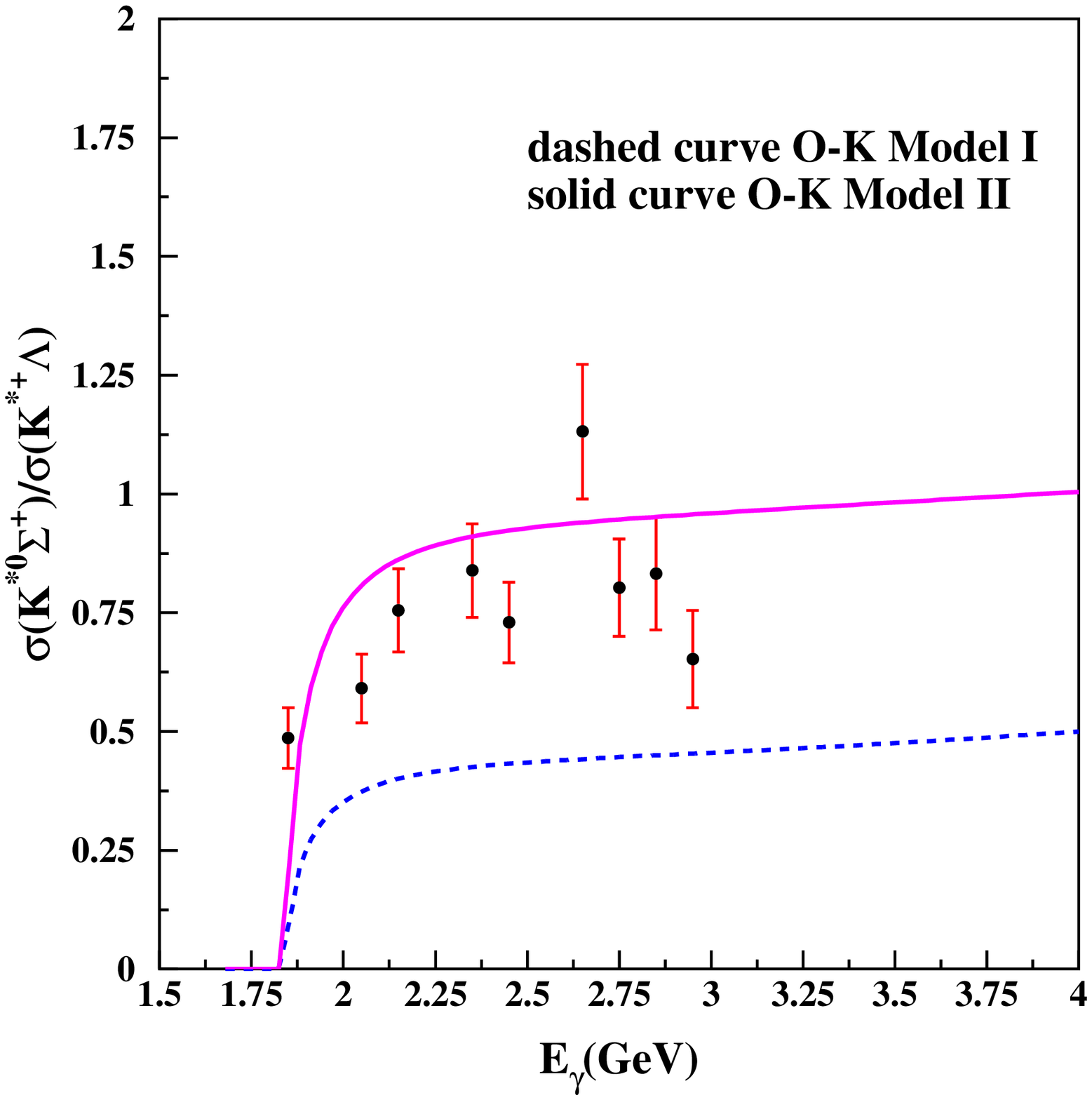,width=\columnwidth}
\caption{Total cross section ratio of the reactions 
$\gamma p \to K^{*0} \Sigma^{+}$ to 
$\gamma p \to K^{*+} \Lambda$. 
The ratio uses the present data in the denominator and data from 
Ref. \cite{hleiqawi} in the numerator.
The dashed and solid curves are theoretical calculations from Oh and Kim 
\cite{oh2} Model I and Model II, respectively. }
\label{ratio}
\end{figure}

Fig.~\ref{ratio} shows the total cross section ratio of the reactions 
$\gamma p \to K^{*0} \Sigma^{+}$ to 
$\gamma p \to K^{*+} \Lambda$. 
The $K^{*+}\Lambda$ data alone are not very sensitive to the $\kappa$ 
exchange due to the unknown strength of the coupling constant, 
$g_{\kappa N \Lambda}$.  However, the coupling constants of these 
two reactions is related in the effective Lagrangian models, and 
so the ratio is sensitive to the effects of $\kappa$ exchange.
The dots with error bars in Fig.~\ref{ratio} 
use the present data along with the previously 
published CLAS data for $K^{*0} \Sigma^{+}$ \cite{hleiqawi}. 
We note that another data set exists for the $K^{*0} \Sigma^+$ 
reaction from CBELSA\cite{cbelsa}, but we have chosen to use 
CLAS data in both numerator and denominator to reduce systematics. 
The two curves are the theoretical predictions from O-K Model 
I and II \cite{oh2}, where Model I includes minimal $t$-channel 
$\kappa$ exchange, while Model II has a significant contribution 
from $\kappa$ exchange.

\section{Discussion and Conclusions}
\label{conclude}

We presented here the first high-statistics measurement of the reactions 
$\gamma p \to K^{*+} \Lambda$ and $\gamma p \to K^{*+} \Sigma^0$. 
The data are from the g11a experiment using the CLAS detector at 
Thomas Jefferson National Accelerator Facility.
Differential cross sections are presented for nine equal-spaced bins in 
$\cos\theta^{CM}_{K^{*+}}$ for each photon energy bin of 0.1 GeV width from 
threshold (1.7 or 1.8 GeV, respectively) up to 3.9 GeV.  
Total cross sections, based on fits to the differential cross sections 
are also presented for both reactions.  

The cross sections for the $K^{*+} \Lambda$ final state are compared 
with calculations from two effective Lagrangian models, one based on 
an isobar model and the other based on the Regge model. Neither 
calculation matches the data over the broad kinematic range measured 
here, but the isobar model compares more favorably, especially at 
higher photon energies. However, both models significantly 
underpredict the total cross sections in the range $2.1 < E_\gamma < 3.1$ 
GeV.  Inclusion of all well-known nucleon resonances improves 
agreement with the data in the region of $E_\gamma \sim 2$ GeV, 
but has only a small contribution above $\sim 2.3$ GeV, and cannot 
explain this excess cross sections in the new data.

It remains an open question whether the excess strength of 
the $K^{*+}\Lambda$ final state in this photon energy region is due 
to additional couplings to yet-unidentified nucleon resonances at 
higher mass, or whether it is due to other effects such as 
channel-coupling through final-state interactions or interference 
at the amplitude level with other physics processes such as 
photoproduction of the $K^0 \Sigma^{*+}$ final state.  The 
latter effect was studied using a simplified Monte Carlo generator 
and showed little or no effect due to interference with the 
$K^0 \Sigma^{*+}$ final state, but more sophisticated theoretical 
calculations should be done to study interference effects.

In comparison, the $K^{*+} \Sigma^0$ final state has a sharper 
peak in the total cross section at $W \sim 2.25$ GeV, and falls 
off more quickly with increasing photon energy than for the 
$K^{*+} \Lambda$ final state.  This suggests whatever mechanism 
that causes the excess cross section for the latter final state 
is not present in the $K^{*+} \Sigma^0$ photoproduction.  However, 
theoretical calculations are not yet available for this final 
state, and we must wait for more theoretical development before 
any such conclusion can be reached.  

One of the goals of this measurement was to understand the role of 
the $\kappa$ meson exchange, which can contribute to $K^{*+}$ 
photoproduction but not to $K^+$ photoproduction.  Although no 
definite conclusion can be reached from the present data, the 
ratio of total cross sections for the $K^{*+} \Lambda$ and the 
$K^{*0} \Sigma^+$ final state compared with a similar ratio 
calculated in the model of Oh and Kim suggests that the model 
with significant $\kappa$ exchange is in better agreement with the 
data ratio.  This agrees with the conclusion from a recent study 
of the beam asymmetry measurement \cite{hwang} of the $K^{*0} \Sigma^+$ 
final state using a linearly polarized photon beam at forward angles.  
However, we must be careful in making any firm conclusion regarding the 
role of the $\kappa$ exchange until the theoretical models have 
better agreement with the $K^{*+} \Lambda$ total cross sections 
above $\sim 2.1$ GeV.  The excess strength of the new data above 
2.1 GeV may change the effects of $\kappa$ exchange in the 
ratio.  However, the general idea of comparing the $K^{*+} \Lambda$ 
and $K^{*0} \Sigma^+$ cross sections, which are affected differently 
by $\kappa$ exchange, is something that can be studied now that 
these new data are available.

\section{Acknowledgment}
The authors thank the staff of the Thomas
Jefferson National Accelerator Facility who made this experiment possible.
This work was supported in part by 
the Chilean Comisi\'on Nacional de Investigaci\'on Cient\'ifica y Tecnol\'ogica (CONICYT),
 the Italian Istituto Nazionale di Fisica Nucleare,
the French Centre National de la Recherche Scientifique,
the French Commissariat \`{a} l'Energie Atomique,
the U.S. Department of Energy,
the National Science Foundation,
the UK Science and Technology Facilities Council (STFC),
the Scottish Universities Physics Alliance (SUPA),
and the National Research Foundation of Korea.

The Southeastern Universities Research Association (SURA) operates the
Thomas Jefferson National Accelerator Facility for the United States
Department of Energy under contract DE-AC05-84ER40150.

\end{document}